\documentclass[sigconf, nonacm]{acmart}
\AtBeginDocument{%
  }

\usepackage{subcaption} % 用于子图的标题  
\usepackage{xcolor}  % 用于自定义color
\usepackage{tcolorbox}  
\usepackage{multirow} % 用于表格
\usepackage[normalem]{ulem}
\usepackage{float}
\usepackage{placeins}

\definecolor{defaultpurple}{rgb}{0.4, 0.031, 0.455} % 转换后的 RGB 值（/255） 
\definecolor{defaultgrey}{rgb}{0.5529, 0.5686, 0.5804} % 转换后的 RGB 值（/255） 
\definecolor{defaultgreen}{rgb}{0.5294, 0.4588, 0.3020} % 转换后的 RGB 值（/255）

\newtcbox{\mybox}[1][defaultpurple]{
  on line,
  arc=1pt,
  colback=white!75!#1,   % 90% 白色 + 10% 原色
  colframe=white!25!#1, 
  fontupper={\color{black}},
  boxsep=0pt,
  left=1.5pt,
  right=1.5pt,
  top=1pt,
  bottom=1pt,
  boxrule=0.5pt       % 边框宽度
}
\begin{document}

\title[PACEE]{PACEE: Parent-Centered AI Scaffolding for Emotion Education in Early Childhood Conversations}
\author{Yu Mei}
\affiliation{%
  % \department{Department of Computer Science and Technology}
  \institution{Tsinghua University}
  \city{Beijing}
  \country{China}
}
\email{meiy24@mails.tsinghua.edu.cn}

\author{Xutong Wang}
\affiliation{%
  % \department{Department of Computer Science and Technology}
  \institution{Tsinghua University}
  \city{Beijing}
  \country{China}
}
\email{wangxuto23@mails.tsinghua.edu.cn}

\author{Ziyao Zhang}
\affiliation{%
  % \department{Department of Computer Science and Technology}
  \institution{Tsinghua University}
  \city{Beijing}
  \country{China}
}
\email{ziyao-zh24@mails.tsinghua.edu.cn}

\author{Yiming Fu}
\affiliation{%
  \institution{Dublin City University}
  \city{Dublin}
  \country{Ireland}
}
\email{yiming.fu3@mail.dcu.ie}

\author{Shiyi Wang}
\affiliation{%
  % \department{Academy of Arts \& Design}
  \institution{Tsinghua University}
  \city{Beijing}
  \country{China}
}
\email{shiyi-wa23@mails.tsinghua.edu.cn}

\author{Qingyang Wan}
\affiliation{%
  % \department{Academy of Arts \& Design}
  \institution{Tsinghua University}
  \city{Beijing}
  \country{China}
}
\email{wanqy23@mails.tsinghua.edu.cn}

\author{Qinghuan Lan}
\affiliation{%
  % \department{Department of Computer Science and Technology}
  \institution{Tsinghua University}
  \city{Beijing}
  \country{China}
}
\email{lqh24@mail.tsinghua.edu.cn}

\author{Chang Liu}
\affiliation{%
  % \department{Department of Computer Science and Technology}
  \institution{Tsinghua University}
  \city{Beijing}
  \country{China}
}
\email{c-liu21@tsinghua.org.cn}

\author{Jie Cai}
\affiliation{%
  % \department{Department of Computer Science and Technology}
  \institution{Tsinghua University}
  \city{Beijing}
  \country{China}
}
\email{jie-cai@mail.tsinghua.edu.cn}

\author{Chun Yu}
\affiliation{%
  % \department{Department of Computer Science and Technology}
  \institution{Tsinghua University}
  \city{Beijing}
  \country{China}
}
\email{chunyu@tsinghua.edu.cn}

\author{Yuanchun Shi}
\affiliation{%
  % \department{Department of Computer Science and Technology}
  \institution{Tsinghua University}
  \city{Beijing}
  \country{China}
}
\email{shiyc@tsinghua.edu.cn}

\renewcommand{\shortauthors}{Mei et al.}

\begin{abstract}
Emotion education is critical for children aged 3 to 6. However, existing technologies largely focus on children’s direct interaction with AI, overlooking the central role of parents in guiding early emotional development at home. To address this gap, we conducted co-design sessions with five kindergarten teachers and five parents to identify key parental challenges and opportunities for AI support in family emotion education. Based on these insights, we developed PACEE, an LLM-based assistant designed to support parents in guiding children’s emotional development through conversations, rather than directly interacting with children. PACEE provides parent-centered AI scaffolding that supports parents in real-time conversation through personalized guidance, post-hoc reflection through trackable feedback, and understanding children’s emotional states through modeling. We evaluated PACEE with 16 families. Results show that PACEE enhances parent-child engagement, fosters deeper emotional communication, and improves parents’ expertise and overall experience in guiding their children. Our findings extend emotion coaching practices to the context of generative AI and offer design insights for building AI systems that support parent-centered family education.
\end{abstract}

\begin{CCSXML}
<ccs2012>
   <concept>
       <concept_id>10003120.10003121.10003124.10011751</concept_id>
       <concept_desc>Human-centered computing~Collaborative interaction</concept_desc>
       <concept_significance>500</concept_significance>
       </concept>
 </ccs2012>
\end{CCSXML}

\ccsdesc[500]{Human-centered computing~Collaborative interaction}

\keywords{Parent-child Interaction, Emotional Education, Parent-AI Collaboration, LLMs}
\begin{teaserfigure}
\centering
\includegraphics[width=0.9\textwidth]{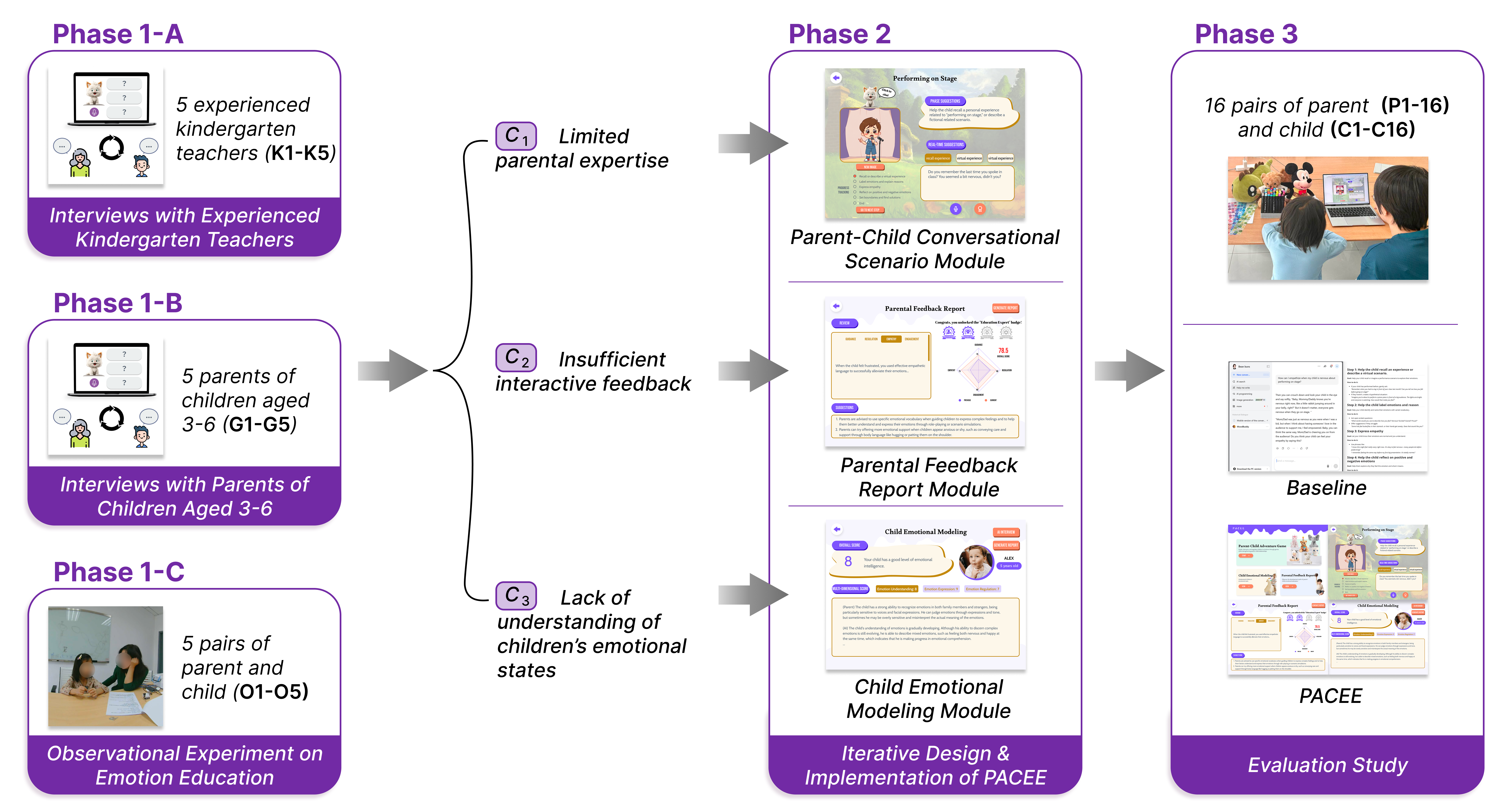}
\caption{Our 3-phased study procedure.}
\label{figure:overall-study-procedure}
\end{teaserfigure}


\maketitle

\section{INTRODUCTION}
Emotion education is critical for children aged 3 to 6. During this period, children begin to understand emotions such as anger, fear, and sadness, and learn to express and regulate these feelings \cite{bartlett1982emotional, saarni1999development, wulandari2021development}.

Recent work has explored AI systems to support children's emotional development, enabling them to recognize, express, and manage emotions. However, most systems target children aged 8 and above, and rely on direct interaction with LLM-based systems \cite{seo2024chacha, tang2024emoeden, shen2025easel}. Such child-facing designs are less suitable for younger children (ages 3–6) for two reasons: (1) they experience emotions intensely but have limited ability to verbalize them \cite{sun2024exploring}, and (2) they rely heavily on parental guidance in early emotional development, as parents mediate both technology use and provide emotional security and co-regulation at this stage \cite{fadlillah2022analysis, alegre2011parenting, toran2024parent, morawska2023managing, corwin2012secure, brock2016interparental}.

Despite this central role, effective emotion education remains challenging for parents. Many lack expertise in child psychology and communication strategies \cite{sanders1999triple, saarni1999development, cole1994development}. In practice, parents often struggle to interpret children's ambiguous emotional signals \cite{castro2015parents} and to support emotional regulation through conversation \cite{meng2020effects}.

To better understand these challenges, we conducted interviews with five kindergarten teachers and five parents. Our findings reveal three key challenges: limited parental expertise, insufficient feedback, and a lack of understanding of children's emotional states. Based on these insights, we developed PACEE, an LLM-based assistant designed to support parents, rather than replace them, in guiding children's emotional development. PACEE provides three modules, each addressing one identified challenge.

We evaluated PACEE's performance through a user study with 16 parent-child pairs. Compared to a baseline LLM-assisted approach, PACEE leads to more meaningful parent-child interactions and deeper emotion-related discussions. Children show higher engagement and richer emotional expression, while parents (1) improve their communication confidence and expertise, (2) adopt more effective emotion education strategies, and (3) gain a better understanding of their children's emotional states.

The contributions of this work are as follows:
\begin{itemize}
    \item We identify three key challenges in family emotion education through formative studies with teachers and parents, and derive corresponding design principles.
    \item We present PACEE, a parent-centered AI scaffolding system that supports parents in guiding children's emotional development during conversations.
    \item Through a user study with 16 families, we demonstrate the effectiveness of PACEE in enhancing parent-child emotion education, and provide design implications for parent-centered AI scaffolding in family education.
\end{itemize}
\section{RELATED WORK}

\subsection{Emotion Coaching Techniques and Theories}
Emotion coaching is a parenting technique that supports children's emotional understanding and regulation \cite{fu2022social,havighurst2021tuning}. A widely adopted framework in HCI is Gottman's Five Steps of Emotion Coaching \cite{gable1999heart, gottman2005five}: (1) Be aware of the child's emotion, (2) Recognize the child's expression of emotion as a perfect moment for intimacy and teaching, (3) Listen with empathy and validate the child's feelings, (4) Help the child learn to label their emotions with words, and (5) Set limits when helping the child to solve problems.

This guideline is often complemented by principles from positive psychology \cite{baker2017positive, fan2018emostory}, such as the PERMA model \cite{seligman2010flourish}. The PERMA model characterizes well-being through Positive Emotion, Engagement, Relationships, Meaning, and Achievement, and has been applied in HCI areas such as well-being-centered design \cite{kovich2023application}, persuasive technologies \cite{nebrida2018m}, and digital mental health \cite{chisale2022perma}.

Existing HCI systems that apply emotion coaching primarily target older children or clinical populations. Prior work explores AI-supported emotional expression and feedback for school-aged children and adolescents \cite{seo2024chacha, kim2018can, ibrahim2025designing}, as well as specialized support for conditions such as autism \cite{tang2024emoeden, ghiotti2023prototyping, sharma2016promoting} or trauma \cite{grassl2024coding}. However, little attention has been given to children aged 3–6.

Compared to older children, 3- to 6-year-olds experience emotions more situationally and intensely but have limited verbalization and self-regulation abilities \cite{sun2024exploring}. These traits make parental guidance central to the development of emotional competence \cite{gottman1996parental, hightower2019exploring}. Yet many parents face barriers, such as limited time and insufficient knowledge of emotion guidance \cite{ahmadpour2023understanding, quan2025parents, BROWN2022126}. Existing tools often assume children can interact with AI independently and fail to empower parents in their coaching role \cite{sharma2025robot}. To address this gap, we propose a parent-centered, theory-grounded approach that supports parents as primary emotion coaches for children aged 3–6.

\subsection{Parent-AI Collaborative Coaching System}
Recent work has explored how AI supports children's emotional and cognitive development in family contexts. Existing systems generally fall into three categories based on AI's role in parent–child interaction: AI-centered, parent-centered, and parent–AI collaborative approaches.

AI-centered systems position AI as a teaching agent that interacts directly with children, providing personalized learning content \cite{tang2024emoeden}, story-based learning \cite{xu2023mathkingdom, zhang2024mathemyths}, and prompts to support language expression \cite{seo2024chacha}. Parent-centered systems treat parents as primary educators, with AI as a support tool without direct child interaction. Prior studies have examined parents' needs and practices through interviews and co-design \cite{dumaru2023after, cingel2017parents, nikkhah2021helping, hiniker2016not, mazmanian2017okay, wang2021protection}, identifying opportunities such as media co-use \cite{yu2024parent} and reflective prompts for parents \cite{ibrahim2025designing}, though many stop at design insights with limited system-level support. Parent–AI collaborative systems combine both roles through task division: parents often act as facilitators or evaluators \cite{slovak2016scaffolding, chen2020new, yarosh2013almost}, while AI supports storytelling, communication, and structured activities \cite{choi2024aacesstalk, zhang2022storybuddy, jonas2023supporting, xiao2012supporting}. 

We extend the latter two lines of work to early childhood contexts by proposing a parent-centered, AI-supported coaching approach that strengthens parents' role in everyday emotion guidance.

\subsection{Modeling Children's Emotional States}
Personalized instruction is important for education and children's emotional development \cite{bernacki2021systematic, dewey2024democracy}. Effective personalization requires modeling children's emotional states. Prior work addresses this through content frameworks and assessment methods.

For content design, the Test of Emotion Comprehension (TEC) \cite{pons2004emotion,albanese2006children} is widely used to assess children's emotional competence with nine key elements, including emotion recognition emotion recognition \cite{bullock1985further}, emotional expression \cite{barden1980children}, emotion external cause understanding \cite{harris1989young}, emotion-belief link \cite{fonagy1997relationship}, emotion reminder effect \cite{harris1983children}, emotional masking \cite{gardner1988japanese}, mixed emotional understanding \cite{arsenio1995children}, emotional morality \cite{harter1989developmental}, and emotion regulation \cite{altshuler1989developmental}.

For assessment, traditional approaches rely on parent- or teacher-reported questionnaires \cite{denham1997parental,rothbart2004temperament}, while self-report tools such as TEIQue-CF \cite{mavroveli2008investigation} capture children's self-perceived competence. However, for children aged 3–6, limited cognitive and expressive abilities can reduce reliability \cite{wellman2004scaling}. More recently, AI-based systems using chatbots or games enable automated observation \cite{santos2020therapist,nicolaidou2022gamified}, but often struggle to interpret subtle developmental cues \cite{calvo2010affect}.

To address these limitations, we propose a human–AI collaborative framework that combines parents' real-world understanding with AI-based analysis to support children's emotional modeling.
\section{STUDY PROCEDURE}

This study was approved by the Institutional Review Board, and we obtained consent from all participating parents and children. We conducted three study phases to explore, develop, and evaluate an AI assistant for supporting family emotion education (\autoref{figure:overall-study-procedure}). 

\textbf{Phase 1: Formative study.} To understand needs and challenges in family emotion education, we conducted three formative studies: (1) interviews with 5 experienced kindergarten teachers (\textbf{K1--K5}), (2) interviews with 5 parents of children aged 3-6 (\textbf{G1--G5}), and (3) an observational experiment with 5 parent–child pairs (\textbf{O1--O5}). We identified three key challenges in family emotion education ($C_1$–$C_3$). Participants were recruited via social media, and each received a \$10 stipend. Detailed demographics are provided in \autoref{appendix:Additional Formative Study Details}. 

\textbf{Phase 2: System design and implementation.} To mitigate these challenges, we developed three core modules of PACEE: Parent-Child Conversational Scenario ($C_1$), Parental Feedback Report ($C_2$), and Child Emotional Modeling ($C_3$). PACEE supports parents in operationalizing Gottman's Five Steps of Emotion Coaching \cite{gable1999heart} and incorporates design insights from the formative studies. 

\textbf{Phase 3: Evaluation study.} We conducted a within-subject study with 16 parent–child pairs to evaluate PACEE, using a generic LLM chatbot as the baseline. 

\section{FORMATIVE STUDY}

% We conducted three formative studies to understand needs and challenges in family emotion education: (1) interviews with 5 experienced kindergarten teachers (\textbf{K1--K5}), (2) interviews with 5 parents of children aged 3-6 (\textbf{G1--G5}), and (3) an observational experiment with 5 parent–child pairs (\textbf{O1--O5}). Participants were recruited via social media and each received a \$10 stipend. Detailed demographics are provided in Appendix~\ref{appendix:Additional Formative Study Details}.

\subsection{Procedure and Analysis}
\subsubsection{Need-finding Interview with Experienced Kindergarten Teachers}
\begin{figure}[h]  
    \centering  
    \begin{subfigure}[b]{0.65\columnwidth}  
        \includegraphics[width=\columnwidth]{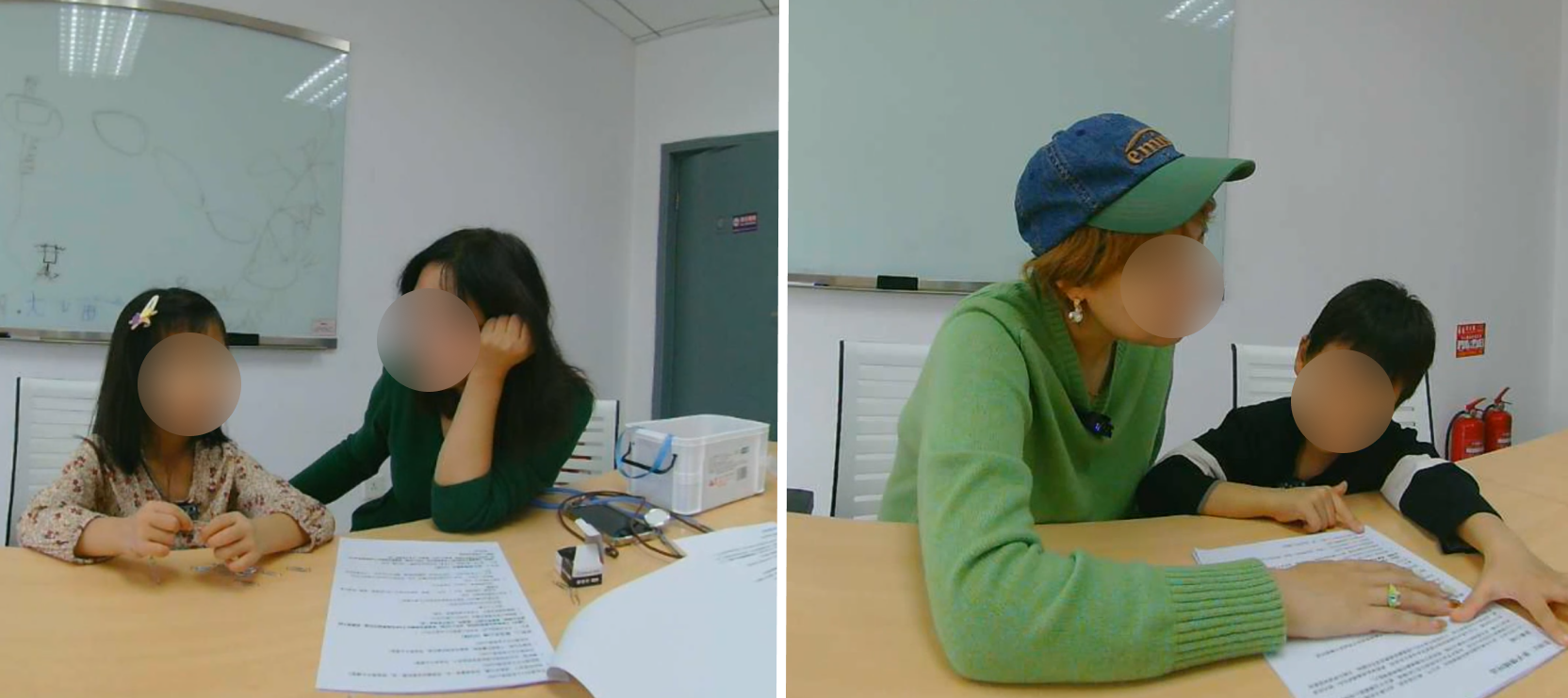} 
        \caption{}  
        \label{fig:formative_study_photoshot}  
    \end{subfigure}  
    \begin{subfigure}[b]{0.25\columnwidth}  
        \includegraphics[width=\columnwidth]{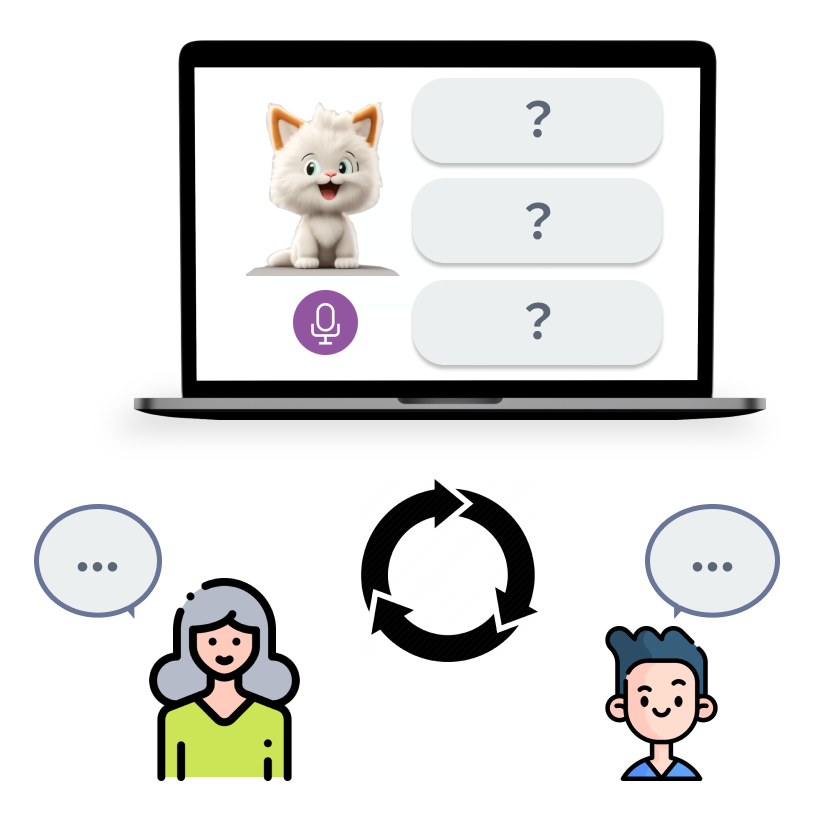} 
        \caption{}  
        \label{fig:formative_study_conceptual_design_images}  
    \end{subfigure}  
    \caption{Formative study scenes: (a) observational experiment setup; (b) conceptual design image used in interviews.}
    \label{fig:formative_study_images}  
\end{figure}

We interviewed 5 kindergarten teachers (3 female, 2 male; $M=9.40$ years teaching experience, $SD=6.34$) in one-hour video conference sessions. 
We discussed common negative emotions among children aged 3-6, their triggers, and teachers' guidance strategies, and explored how AI might support family emotion education. Using conceptual design images (\autoref{fig:formative_study_conceptual_design_images}), we co-designed potential AI-assisted interaction scenarios where parents guide children's emotions at home with AI support.

\subsubsection{Need-finding Interview with Parents of Children Aged 3-6}
Using the same conceptual design images, we interviewed 5 parents (4 mothers, 1 father; $M=41.0$ years old, $SD=3.16$) of children aged 3-6 in one-hour video conference sessions. 
Parents described their children's emotional behaviors, challenges in emotion guidance, and expectations for AI-assisted tools.

\subsubsection{Observational Experiment on Children's Emotion Education}
To understand parent–child interactions during emotion education, we recruited and observed 5 families (parents: 4 mothers, 1 father; $M=38.2$ years old; children: 4 boys, 1 girl; $M=4.8$ years old). In a 30-minute lab session, parents guided children through one of three negative emotion scenarios following Gottman's Five Steps of Emotion Coaching \cite{gable1999heart}. Sessions were video-recorded, followed by short interviews about parents' experiences and difficulties.

All interview and observation data were transcribed and analyzed using thematic coding by three HCI researchers \cite{clarke2017thematic}.

\subsection{Findings}

\subsubsection{Key Challenges in Family Emotion Education}

Through interviews with 5 kindergarten teachers (K1-K5) and 5 parents (G1-G5), we identified common scenarios where children aged 3–6 struggle with negative emotions that require repeated guidance. The findings serve as the initial entries for our system's scenario database (see \autoref{appendix:Additional Formative Study Details}). Parental interviews and observational studies further revealed three key challenges.

\mybox[defaultpurple]{$C_1$} \textbf{. Limited parental expertise.} All interviewed parents (5/5) reported lacking professional knowledge and practical strategies for guiding children's emotions. Parents struggled to translate general theoretical guidelines into concrete actions (G2: ``\textit{the examples in books don't always apply to my kid's situation}''), and observational data showed that they often ``\textit{don't know what to say}'' in certain steps in the guideline (O1). Parents also emphasized the need for ``\textit{concrete, easy-to-implement language and techniques}'' (G1, G2, G5).

\mybox[defaultpurple]{$C_2$} \textbf{. Insufficient feedback.} Parents often lack objective feedback when guiding children's emotional development. As G3 noted, ``\textit{When he has strong emotions, I explain and comfort him in the moment, but I can't tell if he really understood or whether my approach was effective}''. A similar uncertainty appeared in the observational study (O4), where parents followed the guidance steps but remained unsure whether the approach worked.

\mybox[defaultpurple]{$C_3$} \textbf{. A lack of understanding of children's emotional states.} Parents reported difficulty accurately identifying children's emotions and their causes. This is partly due to limited visibility into children's school experiences (G4). Parents may misinterpret emotional triggers (G1) and also acknowledge subjective biases when evaluating their children's emotional development (G5), highlighting the need for more objective assessments.

\subsubsection{Roles and Design Principles of AI in Family Emotion Education}
Interviews with parents explored potential roles of AI in emotional guidance, while kindergarten teachers highlighted developmental characteristics of young children that should inform system design. All parents preferred to retain their primary role in emotional guidance and use AI to complement their limitations. They envisioned three main roles for AI.

First,  AI should act as a \textbf{conversation facilitator}, dynamically adjusting its participation in parent–child interactions (G3:  ``\textit{When parents take the lead, AI should speak less; when parents struggle, AI can step in to enrich the conversation.}''). Second, parents viewed AI as \textbf{an expert resource} that could integrate pedagogical knowledge into conversations and provide objective feedback about children's development. Third, AI could function as \textbf{a generative tool}, producing visual materials, conversation strategies, or performance feedback (G5: ``\textit{It could analyze which guidance techniques I applied successfully and how well I executed them.}'').

Interviews with kindergarten teachers highlighted two design principles for AI-assisted emotion education systems. First, \textbf{systems should align with children's developmental stages}. Given young children's concrete thinking and limited language abilities, designs should incorporate visual aids and provide parents with age-appropriate communication guidance. Second, \textbf{systems should support personalized instruction}. Because emotional and cognitive development varies across children, AI systems should consider individual traits and behavioral patterns to deliver tailored guidance and developmental feedback.

\section{DESIGN AND IMPLEMENTATION OF PACEE}
\label{section: DESIGN AND IMPLEMENTATION OF PACEE}

\begin{figure*}[]
\centering
\includegraphics[width=0.8\textwidth]{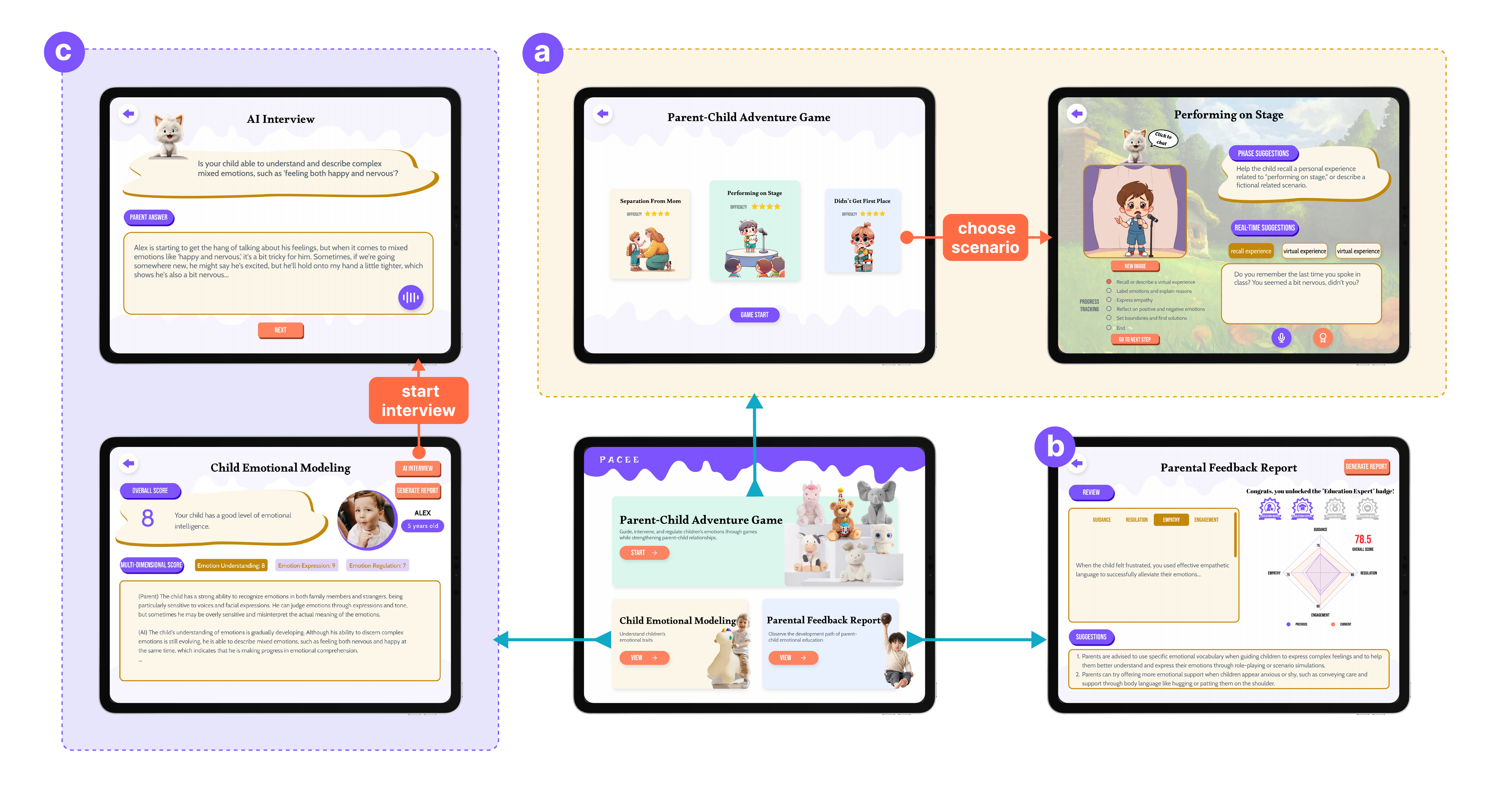}
\caption{Main screens of PACEE showing its three modules: (a) Parent-Child Conversational Scenario, (b) Parental Feedback Report, and (c) Child Emotional Modeling.}
\label{figure:UI}
\end{figure*}

\begin{figure*}[htb] 
\centering
\includegraphics[width=0.85\textwidth]{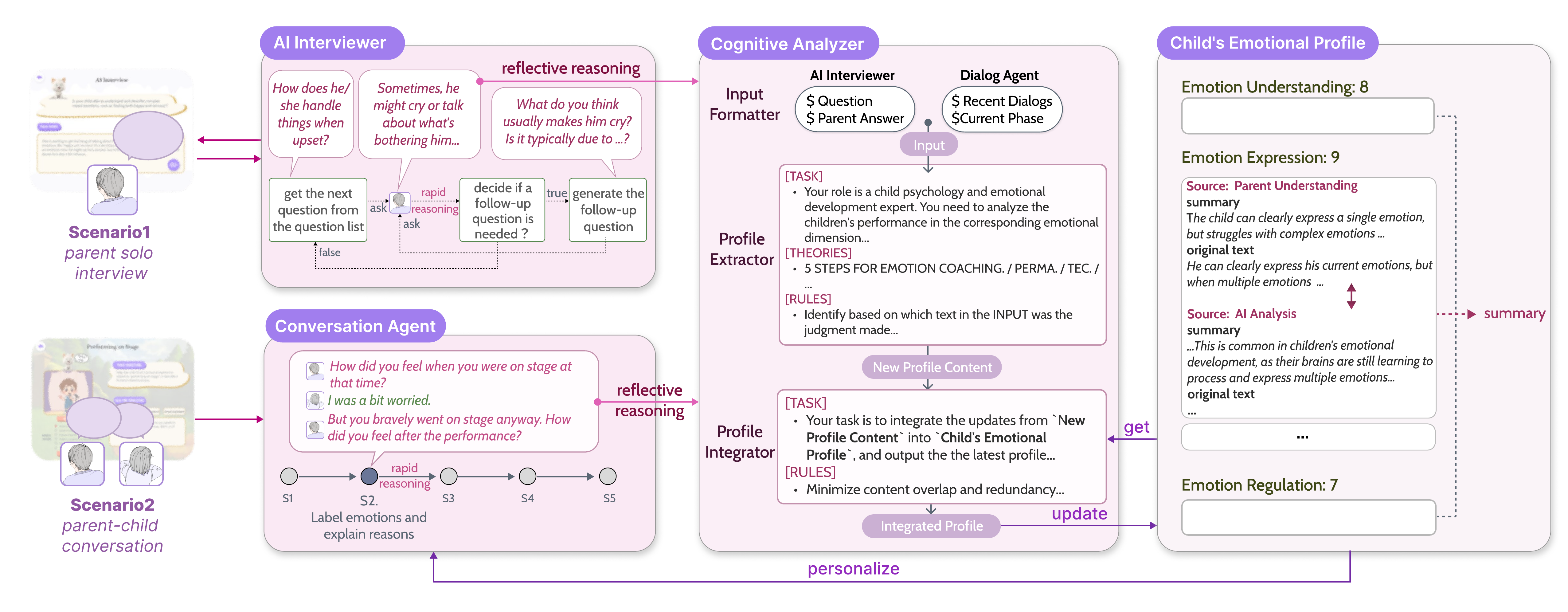}
\caption{PACEE mechanism for collaboratively modeling children's emotional profiles through AI-parent interviews and parent–child conversations.}
\label{figure:modeling_emotional_profiles}
\end{figure*}

\subsection{User Interface and Design of System Modules}
Based on insights from the formative studies, we designed PACEE, an LLM-based assistant that supports parents in guiding emotion-related conversations with their children. Rather than interacting directly with children, PACEE collaborates with parents by providing conversational support, feedback, and emotional profiling. The system contains three modules: \textit{Parent-Child Conversational Scenario}, \textit{Parental Feedback Report}, and \textit{Child Emotional Modeling}, each addressing a key challenge in family emotion education.

\subsubsection{Parent-Child Conversational Scenario}
To address \mybox[defaultpurple]{$C_1$} limited parental expertise, the \textit{Parent-Child Conversational Scenario} module supports parents in conducting scenario-based conversations with their children. Parents guide conversations about their children's emotions, while PACEE provides real-time assistance. Parents begin by selecting a situation (e.g., performing on stage) from a scenario database derived from the formative study. 

PACEE supports parents in operationalizing Gottman's Five Steps of Emotion Coaching \cite{gable1999heart} as a five-stage conversational workflow (\autoref{tab:5-steps-mapping}). In this workflow, PACEE does not directly coach children; instead, it provides real-time support to parents across the five stages \mybox[defaultgrey]{$S_1$}-\mybox[defaultgrey]{$S_5$}. Specifically, PACEE facilitates conversation flow, provides stage-aware guidance, generates scenario images to help parents make situations more concrete, allows parents to optionally deliver AI-generated utterances when they need support in expressing their ideas more effectively, and tracks dialogue progress to support parents' awareness of the conversation. 

\begin{table}[]
\caption{PACEE operationalizes Gottman's Five Steps of Emotion Coaching \cite{gable1999heart} to supports parents interacting children aged 3-6.}
\resizebox{\columnwidth}{!}{
\begin{tabular}{llll}
\toprule[1pt]
\multicolumn{1}{c}{\textbf{Tag}} & \multicolumn{1}{c}{\textbf{Gottman's Guideline \cite{gable1999heart}}}                                              & \multicolumn{1}{c}{\textbf{\begin{tabular}[c]{@{}c@{}}Developmental Consideration \\ of Children Aged 3-6\end{tabular}}}                                        & \multicolumn{1}{c}{\textbf{PACEE Support for Parents}}                                                          \\ \hline
\mybox[defaultgrey]{$S_1$}       & \begin{tabular}[c]{@{}l@{}}Be aware of the child's \\ emotion. Recognize the \\ expression of emotions\end{tabular} & \begin{tabular}[c]{@{}l@{}}Have limited memory recall.\\ Need for concrete context.\end{tabular}                                                      & \begin{tabular}[c]{@{}l@{}}Help the child recall an \\ experience or describe a \\ virtual scenario\end{tabular} \\ \hline
\mybox[defaultgrey]{$S_2$}       & \begin{tabular}[c]{@{}l@{}}Help the child learn to \\ label their emotions\end{tabular}                             & \begin{tabular}[c]{@{}l@{}}Need scaffolding to articulate \\ feelings\end{tabular}                                                                    & \begin{tabular}[c]{@{}l@{}}Help the child label \\ emotions and reason\end{tabular}                              \\ \hline
\mybox[defaultgrey]{$S_3$}       & Listen with empathy                                                                                                 & \multirow{2}{*}{\begin{tabular}[c]{@{}l@{}}Respond well to positive \\ reinforcement (e.g., PERMA \\ model \cite{seligman2010flourish})\end{tabular}} & Express empathy to children                                                                                      \\ \cline{1-2} \cline{4-4} 
\mybox[defaultgrey]{$S_4$}       & \begin{tabular}[c]{@{}l@{}}Validate the child's \\ feelings\end{tabular}                                            &                                                                                                                                                       & \begin{tabular}[c]{@{}l@{}}Help the child reflect on \\ positive and negative emotions\end{tabular}              \\ \hline
\mybox[defaultgrey]{$S_5$}       & \begin{tabular}[c]{@{}l@{}}Set limits when helping \\ the child to solve problem\end{tabular}                       & \begin{tabular}[c]{@{}l@{}}Need help or collaboration\\ in problem solving\end{tabular}                                                               & \begin{tabular}[c]{@{}l@{}}Help the child set boundaries \\ and find positive solutions\end{tabular}             \\ \bottomrule[1pt]
\end{tabular}
}
\label{tab:5-steps-mapping}
\end{table}

\subsubsection{Parental Feedback Report}
To address \mybox[defaultpurple]{$C_2$} insufficient feedback, the \textit{Parental Feedback Report} evaluates parental guidance during the \textit{Parent-Child Conversational Scenario} and provides actionable suggestions for future interactions.

PACEE analyzes transcripts of parent-child conversations and generates personalized recommendations grounded in established emotional education frameworks (e.g., Gottman's guideline \cite{gable1999heart}) and early childhood development guidelines for ages 3–6 \cite{MOE_3to6_ELDG_2012}. The feedback evaluates how parents performed across the five stages of emotion coaching (\mybox[defaultgrey]{$S_1$}-\mybox[defaultgrey]{$S_5$}) and provides suggestions to improve their future guidance at each stage.

\subsubsection{Child Emotional Modeling}
\label{section: Child Emotional Modeling}
To address \mybox[defaultpurple]{$C_3$} limited understanding of children's emotional states, the \textit{Child Emotional Modeling} module constructs emotional profiles informed by the TEC framework \cite{pons2004emotion,albanese2006children}. This module supports \mybox[defaultgrey]{$S_1$} by helping parents recognize their children's emotional states and developmental characteristics, while also providing cues that facilitate empathetic listening in \mybox[defaultgrey]{$S_3$}. The model integrates two sources: (1) parents' long-term observations collected through a semi-structured AI interview, and (2) behavioral analysis of parent–child conversations during the \textit{Parent-Child Conversational Scenario} (\autoref{figure:modeling_emotional_profiles}).

A cognitive analyzer extracts updates from both sources across three dimensions (emotional understanding, emotional expression, and emotional regulation) and integrates them into the profile. The system also separates parent-reported observations and AI-driven analysis of conversation, allowing parents to compare these sources and identify possible gaps in understanding. The resulting emotional profile is presented on the Child Emotional Modeling interface (\autoref{figure:UI}(c), helping parents better recognize their children's emotional signals in future conversational scenarios.

\subsection{LLM-based Agent Workflow Design}
\begin{figure*}[htb] 
\centering
\includegraphics[width=0.8\textwidth]{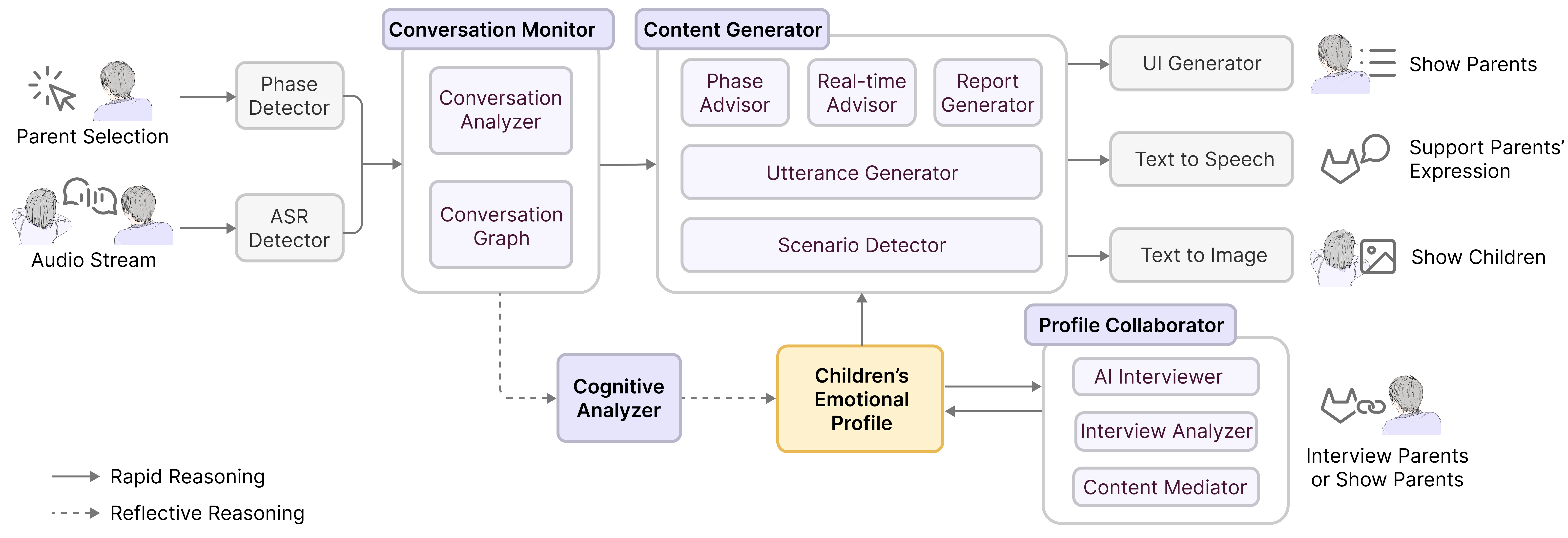}
\caption{Agent workflow of PACEE. }
\label{figure:agent_workflow}
\end{figure*}

We designed an LLM-based agent workflow for PACEE to support real-time parent–child interactions (\autoref{figure:agent_workflow}). During the \textit{Parent-Child Conversational Scenario}, PACEE applies two complementary reasoning loops inspired by ``thinking fast and slow'' \cite{kahneman2011thinking, christakopoulou2024agents, lin2023swiftsage}: (1) rapid reasoning for real-time conversational support and (2) reflective reasoning for modeling the child's emotional characteristics.

PACEE first transcribes parent-child conversations and uses parent-selected stages (\mybox[defaultgrey]{$S_1$}–\mybox[defaultgrey]{$S_5$}) to align interaction data. Within the selected stage, the conversation analyzer estimates real-time interaction progress while maintaining a structured representation of the conversation by organizing utterances and speaker turns.
Based on the conversation context and the child's emotional profile, the content generator applies rapid reasoning to provide real-time support, including stage-specific guidance, real-time suggestions, AI-generated utterances, and scenario images. It also aggregates interaction information to generate the \textit{Parental Feedback Report}.

In parallel, reflective reasoning is applied to update the child's emotional profile from conversation data. The cognitive analyzer periodically performs multi-step reasoning to refine the profile during interactions (see \autoref{section: Child Emotional Modeling}). Implementation details are provided in \autoref{appendix:PACEE System Implementation}.
\section{USER EVALUATION}
\label{section:USER EVALUATION}
We conducted a within-subject study to evaluate PACEE in supporting family emotion education. The study aimed to answer the following questions: 

\textbf{RQ1.} How does PACEE enhance family emotion education?

\textbf{RQ2.} How do parents collaborate with PACEE to develop children's emotional profiles and offer guidance to children?

\textbf{RQ3.} How do parents perceive PACEE in addressing their challenges in family emotion education?

We included a Baseline condition for comparison (\autoref{section: Baseline Condition}). Each parent-child pair completed two emotional scenarios: one in the Baseline condition and the other in the PACEE condition, with scenario-condition assignments counterbalanced across participants.

\subsection{Participants}
We recruited 16 parent-child pairs through social media platforms. Parent participants (P1–16) included four fathers and twelve mothers (aged $M=37.69$, $SD=3.31$). Child participants (C1–16) included seven girls and nine boys aged 3–6 years ($M=5.06$, $SD=0.66$). Detailed demographics are provided in \autoref{appendix:Additional User Evaluation Details}.

The study followed standard ethical procedures. Parents provided informed consent for themselves and their children and could withdraw at any time. During the study, server-side moderation was conducted to review generated text and images, ensuring content safety for children.

\subsection{Baseline Condition}
\label{section: Baseline Condition}

\begin{figure}[htb] 
\centering
\includegraphics[width=0.8\columnwidth]{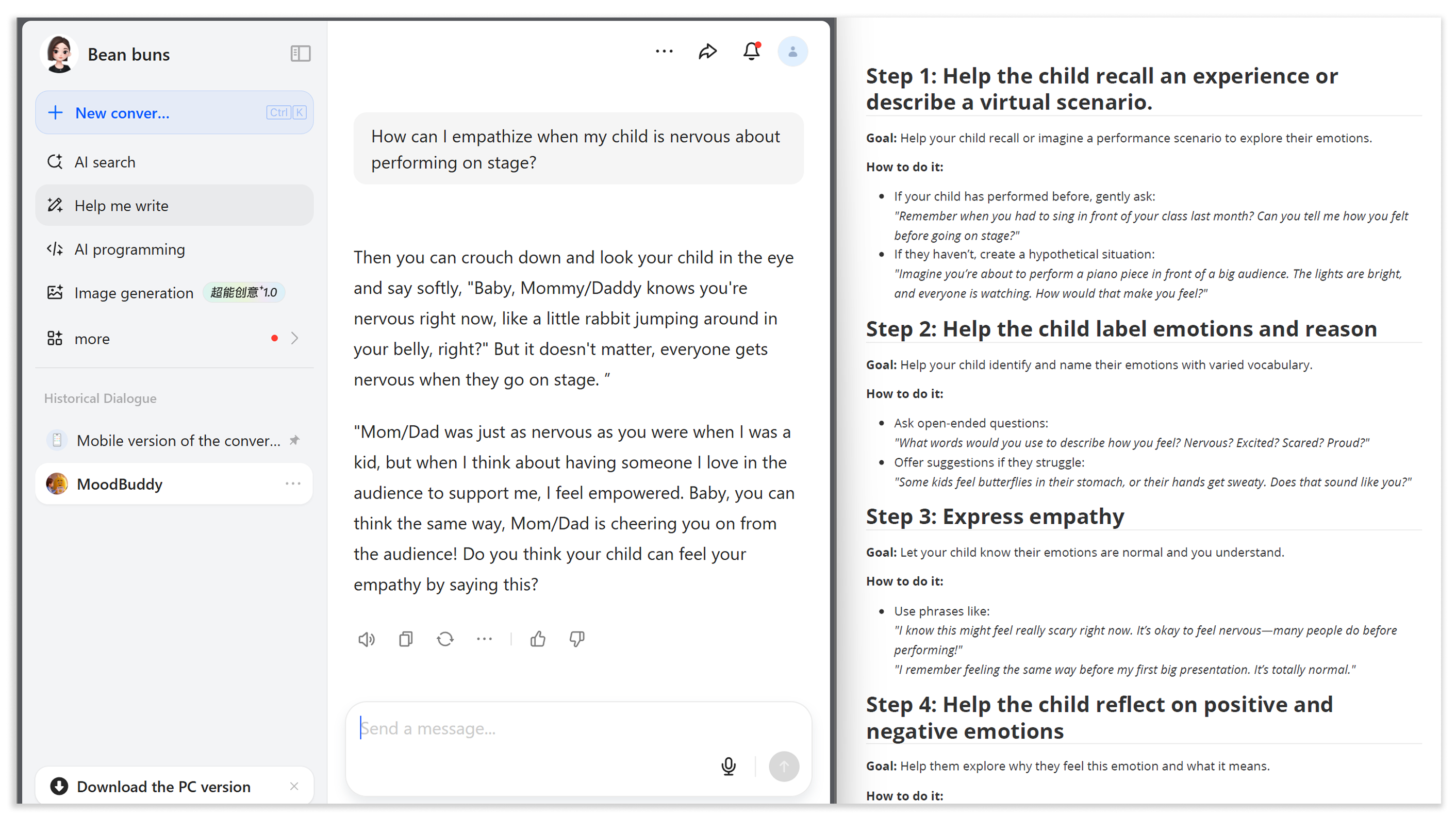}
\caption{Baseline condition. Parents interacted with a voice-based LLM chatbot (Doubao) using the on-screen \mybox[defaultgrey]{$Five\ Stages$} guide, without PACEE features.}
\label{figure:evaluation_study_baseline_condition}
\end{figure}

To evaluate PACEE, we designed a Baseline in which parents interacted with a standard voice-based LLM chatbot (Doubao) through a web interface (\autoref{figure:evaluation_study_baseline_condition}). Parents could request assistance from the chatbot freely during the interaction.

We selected the Doubao chatbot \footnote{\url{https://www.doubao.com/chat/}} as the baseline for three reasons: (1) it is widely used among Chinese families, providing an ecologically valid comparison \cite{newsDoubao1, newsDoubao2, yu2025suitability}; (2) it supports real-time Chinese voice interaction aligned with PACEE's hands-free workflow; and (3) it provides competitive performance on Chinese LLM benchmarks \cite{xu2023superclue}, ensuring the comparison focuses on interaction design rather than model capability.

To ensure structural consistency between conditions, parents were also given an on-screen guide showing the \mybox[defaultgrey]{$Five\ Stages$} conversational workflow. While the stages were fixed, the specific conversational content remained open-ended.

\subsection{Evaluation Design}
We selected two scenarios from the system's database (\autoref{tab:scenario-database}), derived from the formative study: (1) \textit{Not First}, where children may experience anger or disappointment after not winning first place, and (2) \textit{On Stage}, where children may feel anxiety when performing in front of an audience..
We adopted a counterbalanced design. Each pair was randomly assigned to one of four groups in a  $2 \times 2$ design, crossing the order of system conditions (Baseline vs. PACEE) and the assignment of emotional scenarios (\textit{Not First} vs. \textit{On Stage}).

% \begin{figure}[htb] 
% \centering
% \includegraphics[width=0.4\textwidth]{figures/evaluation_study_design.png}
% \caption{Counterbalanced evaluation study design.}
% \label{figure:evaluation_study_design}
% \end{figure}

\subsection{Procedure}
\begin{figure}[h]  
    \centering  
    \begin{subfigure}[b]{0.20\textwidth}  
        \includegraphics[width=\textwidth]{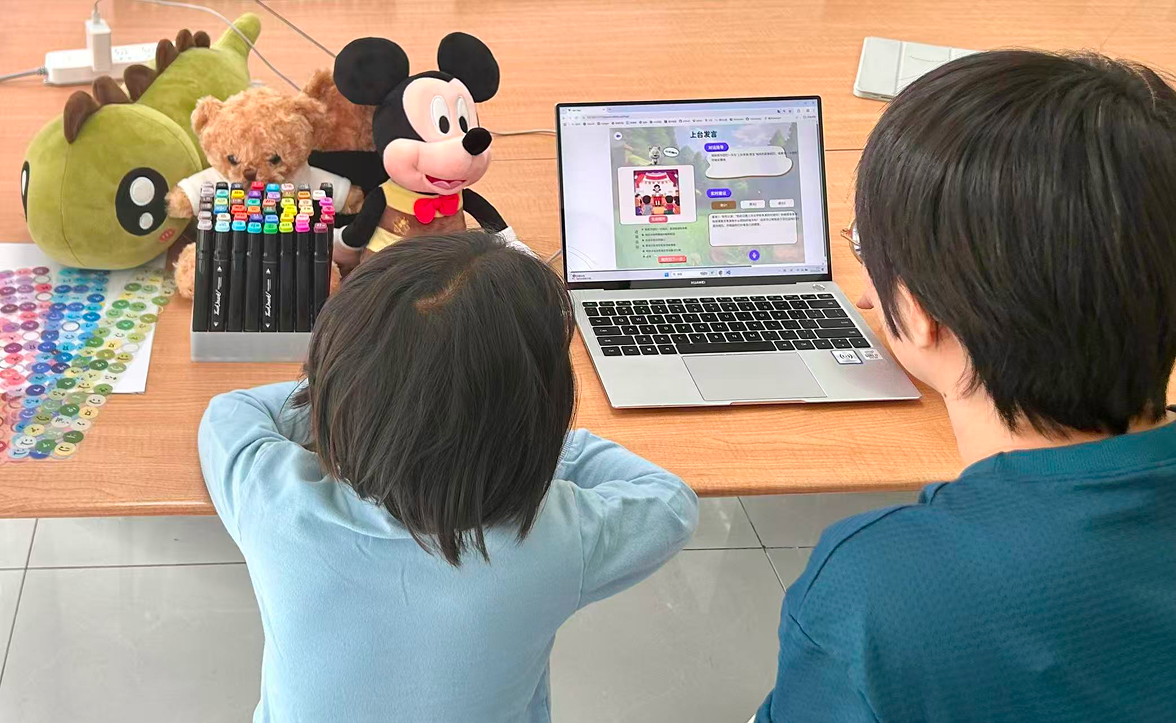} 
        \caption{}  
        \label{fig:evaluation_study_photoshot}  
    \end{subfigure}  
    \hspace{0.04\textwidth} % 调整这里的间距，使其均匀分布
    \begin{subfigure}[b]{0.18\textwidth}  
        \includegraphics[width=\textwidth]{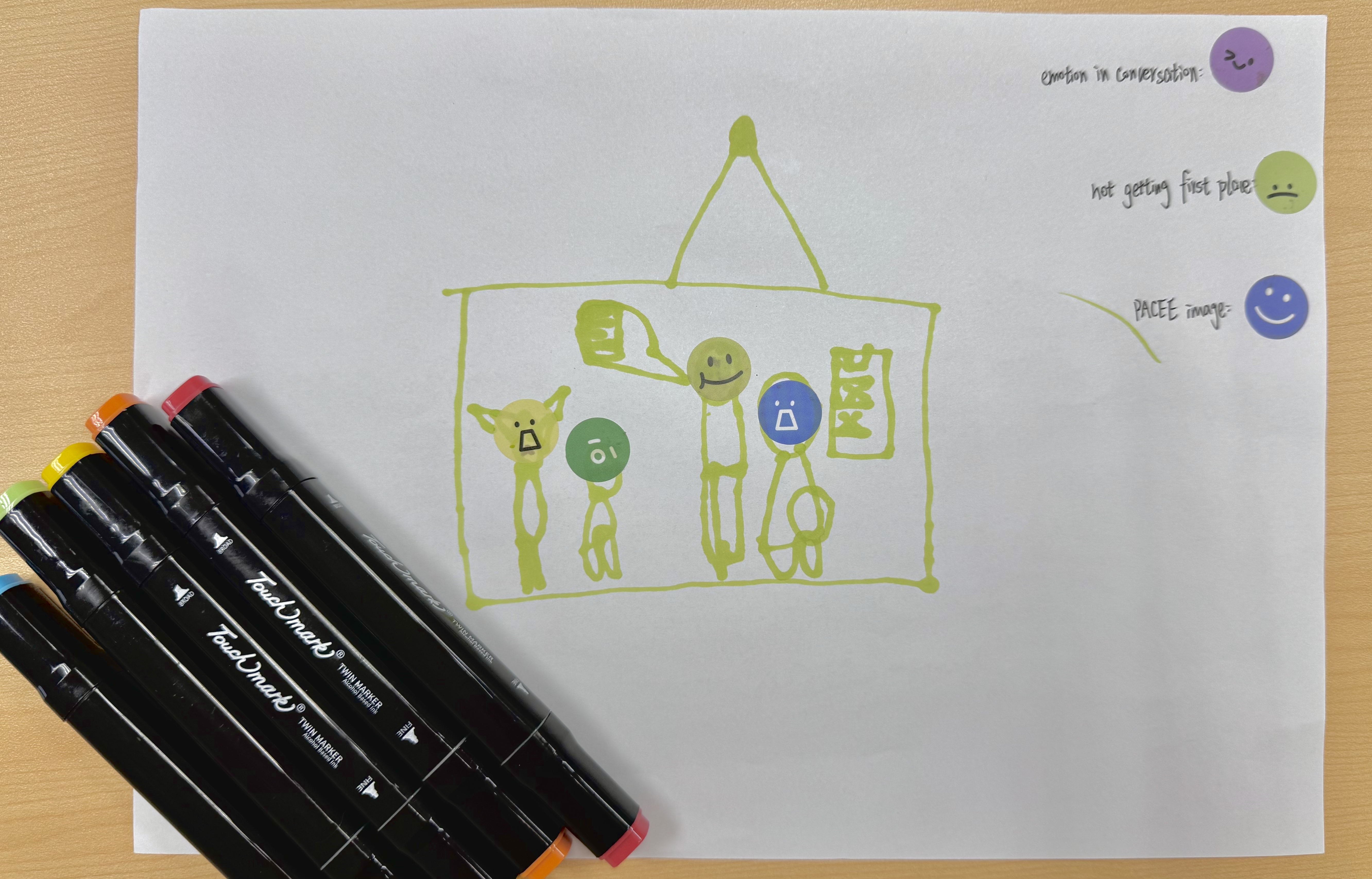} 
        \caption{}  
        \label{fig:evaluation_study_drawing}  
    \end{subfigure}  
    \caption{Scenes from the evaluation study. (a) Parent-child emotional conversation. (b) Child's drawing of the discussed scenario.}  
    \label{fig:evaluation_study_images}  
\end{figure}

Each pair participated in a 2-hour lab study and interacted with the system on a touch-screen laptop (\autoref{fig:evaluation_study_images}). 

\textbf{Introduction.} Parents received a brief system introduction and completed a short practice session. They were instructed to conduct the activity jointly with their child rather than letting the child interact with AI alone.

\textbf{Parent-Child Emotional Conversation.}  Parents used PACEE or the Baseline to conduct emotional conversations with their children based on the assigned scenarios. To reduce potential discomfort for children aged 3–6, the researcher left the room and monitored the session remotely.
    
\textbf{Post-Questionnaire.} Parents completed a questionnaire including NASA-TLX \cite{hart1988development}, the Parenting Sense of Competence Scale (PSOC) \cite{gibaud1978development}, and items assessing how the systems addressed the three challenges (\mybox[defaultpurple]{$C_1$}–\mybox[defaultpurple]{$C_3$}), all rated on 7-point Likert scales.

\textbf{Parent Interview and Child Drawing.} We conducted semi-structured parent interviews, while children expressed their feelings using emoji stickers and drawings of the discussed scenarios.

\subsection{Data Analysis}
\label{section: Data Analysis}
For qualitative analysis, we transcribed all interviews and conducted thematic analysis \cite{braun2006using,terry2017thematic, braun2019reflecting}. Five HCI researchers collaboratively developed a codebook, independently coded the data, and resolved disagreements through discussion. Codes were then clustered to identify recurring themes.

For quantitative analysis, we transcribed the PACEE and Baseline session audio and performed automated text analysis. Emotion-related words were identified using the NTUSD dictionary \cite{ku2007mining}. We computed descriptive statistics and conducted paired-sample t-tests (interaction duration) and Wilcoxon signed-rank tests (dialogue turns, word counts, and emotion-related word counts).

\section{FINDINGS}
\label{section:FINDINGS}

\subsection{RQ1. How Does PACEE Enhance Family Emotion Education?}

\subsubsection{Increased Parent-Child Interaction}

\begin{figure}[htb] 
\centering
\includegraphics[width=0.5\columnwidth]{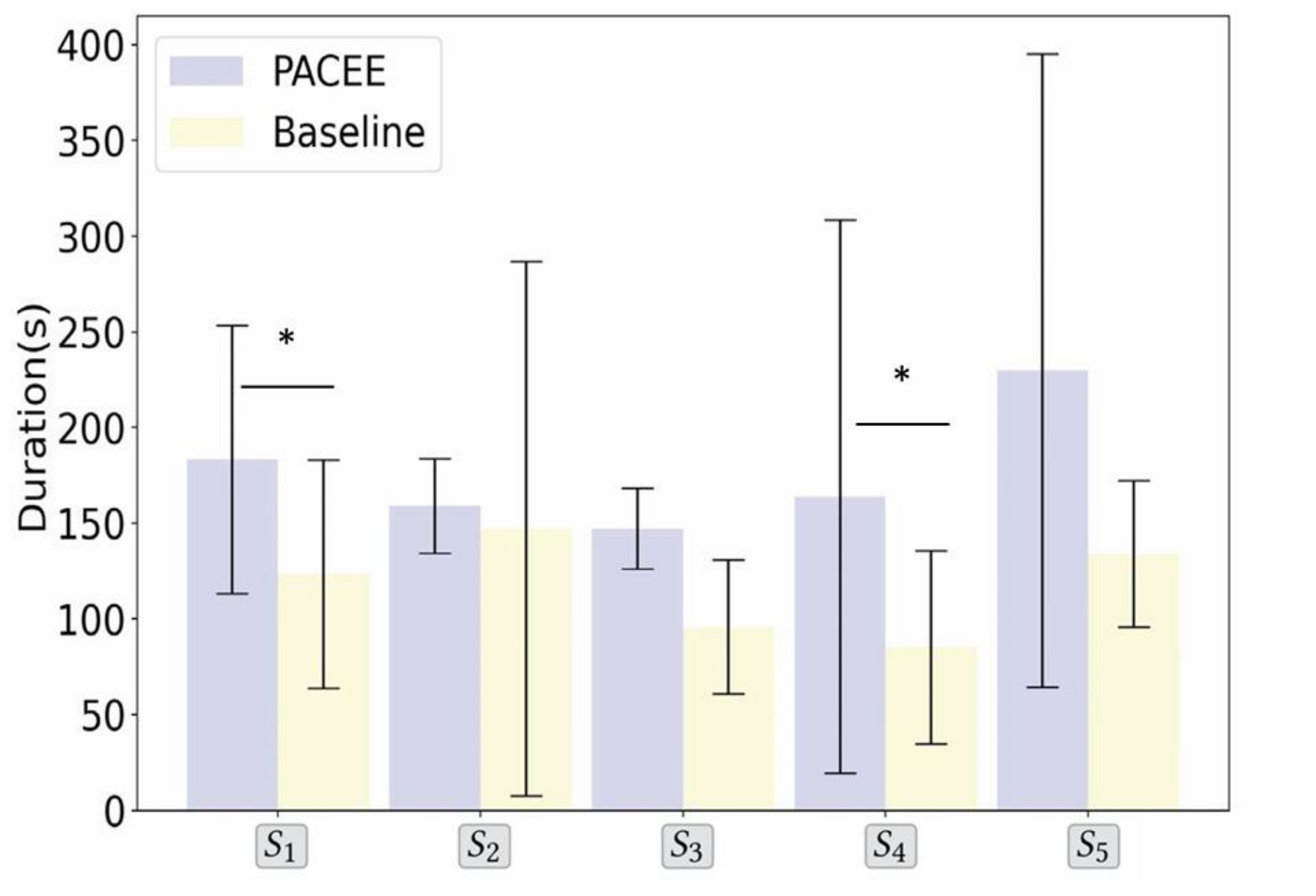}
\caption{User interaction durations in each parent–child conversation stage. Significance results are marked with asterisks (*: p < .05).}
\label{fig:teaching_session_data_5steps}
\end{figure}

To compare interaction outcomes between PACEE and the Baseline, we analyzed interaction duration, dialogue turns, word count, and emotion-related word usage. PACEE outperformed the Baseline across all metrics.

\textbf{Interaction duration.} Participants spent significantly longer time in sessions with PACEE ($M=892.37$, $SD=310.96$ vs. $M=560.17$, $SD=194.61$, $p<0.01$). To further examine how interaction unfolded over time, we analyzed duration across the \mybox[defaultgrey]{$Five\ Stages$} (\autoref{fig:teaching_session_data_5steps}). Participants using PACEE spent more time at each stage, with significant increases in \mybox[defaultgrey]{$S_1$} and \mybox[defaultgrey]{$S_4$} ($p < 0.05$ for both), suggesting that PACEE better supports parents in recalling emotional contexts and reflecting on positive and negative emotions.

\textbf{Dialogue turns and word count.} Participants engaged in more dialogue turns with PACEE ($M=60$, $SD=46.98$ vs. $M=38$, $SD=25.30$, $p<0.05$). Parents produced more words ($M=1597$, $SD=678.56$ vs. $M=994$, $SD=468.36$, $p<0.05$), and children showed a similar increase ($M=370$, $SD=240.43$ vs. $M=165$, $SD=102.25$, $p<0.001$). Given that both systems were matched in workload and participants had flexibility in session completion, these results indicate increased interaction engagement with PACEE.

\textbf{Emotion-related word usage.} Parents used more emotion words with PACEE ($M=111$ vs. $M=70$, $p<0.05$), and children also showed increased emotional expression ($M=19$ vs. $M=10$, $p<0.01$). These findings suggest that PACEE encourages both parents and children to express emotions more openly, fostering richer emotional exchanges.

\subsubsection{Sustained Participation Over Time}
\begin{figure}[htb] 
\centering
\includegraphics[width=\columnwidth]{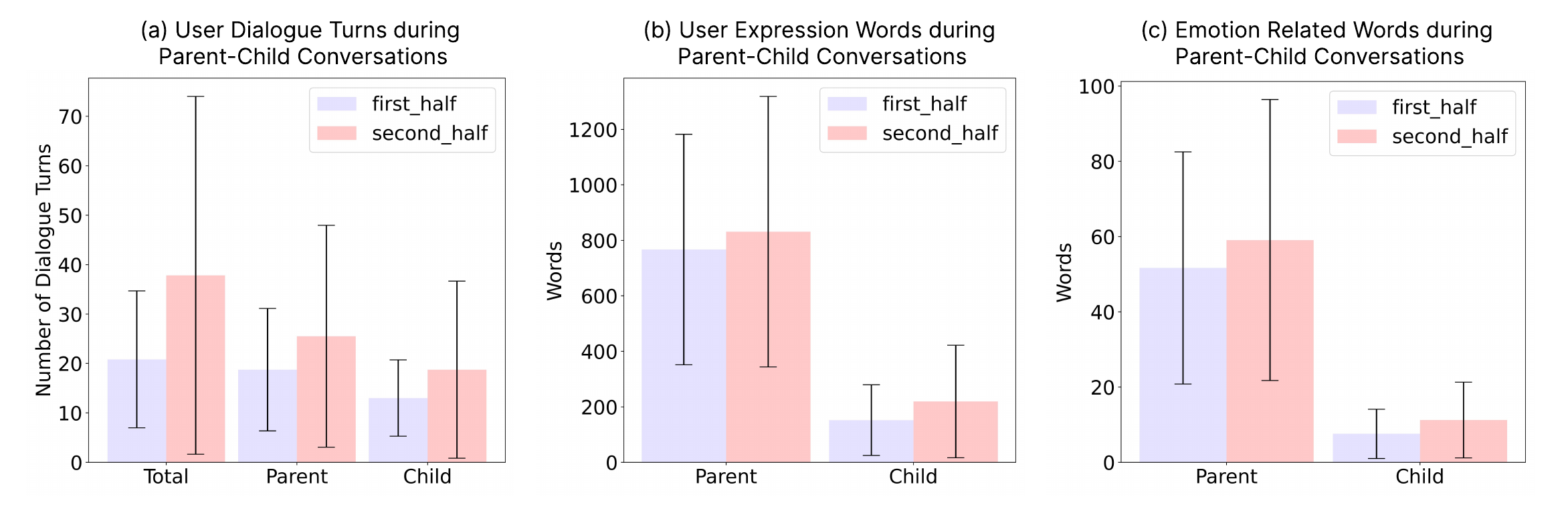}
\caption{Temporal comparison of PACEE interactions between session halves: (a) dialogue turns, (b) word count, and (c) emotion-related words.}
\label{fig:teaching_session_data_halfs}
\end{figure}

To examine potential novelty effects, we divided each PACEE session into two equal temporal halves (\autoref{fig:teaching_session_data_halfs}). No significant differences were found between the first and second halves in dialogue turns, word counts, or emotion-related words. This result suggests that the user's participation in PACEE remained stable throughout the interaction.

\subsubsection{Improved Parental Experience and Children's Engagement Respectively}
We analyzed NASA-TLX and PSOC results from the post-questionnaire. Compared to the Baseline, PACEE significantly improved parents' immediate interaction experience, including higher satisfaction ($p<0.05$, $M=5.06$, $SD=1.12$ vs. $M=3.94$, $SD=1.57$), lower effort ($p<0.01$, $M=2.56$, $SD=1.79$ vs. $M=3.62$, $SD=2.03$), and reduced frustration ($p<0.05$, $M=2.06$, $SD=1.24$ vs. $M=3.06$, $SD=1.57$). In contrast, intrinsic psychological factors (e.g., self-efficacy) showed no significant changes, likely because they require longer-term interventions to produce measurable changes \cite{bandura1997self, sandler2011long}.
% effortAI_PACEE: N=16, M=2.56, SD=1.79
% effortAI_Baseline: N=16, M=3.62, SD=2.03
% frustration_PACEE: N=16, M=2.06, SD=1.24
% frustration_Baseline: N=16, M=3.06, SD=1.57
% satisfaction_PACEE: N=16, M=5.06, SD=1.12
% satisfaction_Baseline: N=16, M=3.94, SD=1.57

We also analyzed children's post-study drawings to understand their perceptions of PACEE. 13 of the 16 children accurately reconstructed the interaction scenes (e.g., stage, parents, and audience) and frequently expressed positive emotions through visual cues such as smiles, hand-holding gestures, and warm color palettes. Additionally, most children (13/16) depicted PACEE as a friendly and supportive figure, often represented as smiling anthropomorphic characters or cartoon animals.

\begin{figure}[htb] 
\centering
\includegraphics[width=\columnwidth]{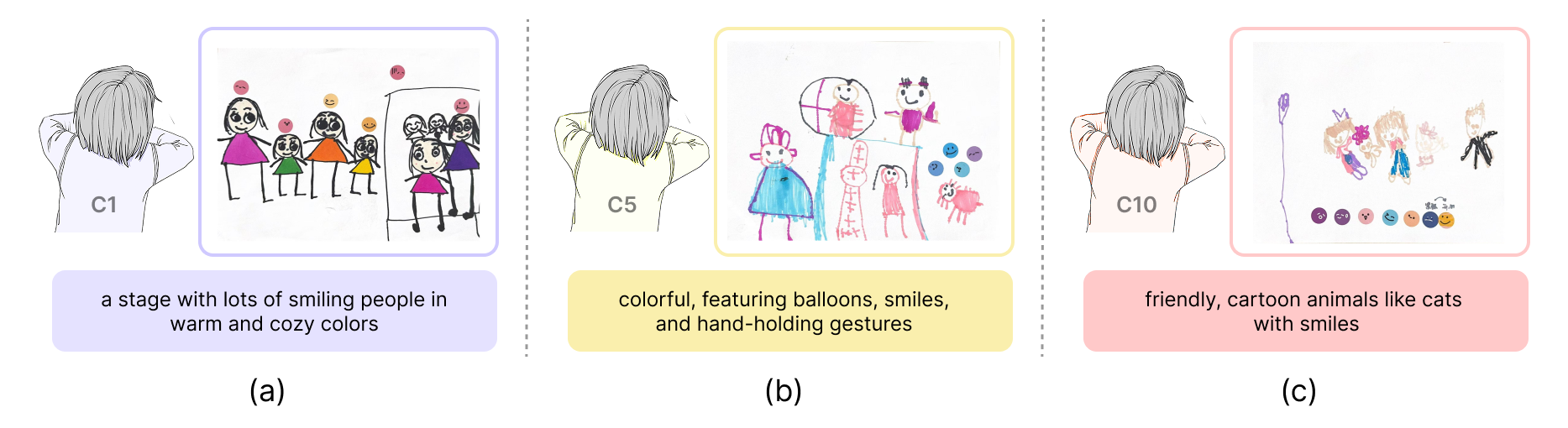}
\caption{Drawings created by children (C1, C5, C10) after parent-child emotional conversations.}
\label{figure:children-drawings}
\end{figure}

\subsection{RQ2. How Do Parents Collaborate with PACEE to Develop Children's Emotional Profiles and Offer Guidance to Children?}

\subsubsection{Parents and PACEE Contribute Different Factors in Construction of Children's Emotional Profiles}
We annotated the system interaction logs using the method described in \autoref{section: Data Analysis}. As shown in \autoref{tab:parent_ai_profile_factors}, the results indicate that parents and PACEE play complementary roles in constructing children's emotional profiles. PACEE primarily derives insights from parent-child interaction data, whereas parents contribute contextual knowledge from firsthand observations, third-party feedback, school progress, and comparisons with how the child behaved at younger ages. In interpreting children's emotions, parents tend to rely on everyday life events and experiential knowledge of behavioral habits, whereas PACEE analyzes interaction patterns to assess children's emotional and developmental characteristics.

\begin{table}[t]
\caption{Parents and PACEE contribute different themes in children's emotional profiles in the evaluation studies.}
\resizebox{\columnwidth}{!}{
\begin{tabular}{llll}
\toprule[1pt]
\multicolumn{1}{c}{\textbf{Source}} & \multicolumn{1}{c}{\textbf{Theme}}                                        & \multicolumn{1}{c}{\textbf{Sub-theme (Partial)}}                                                                          & \multicolumn{1}{c}{\textbf{Example}}                                                                                                                                                                                                                                                                 \\ \hline
\multirow{4}{*}{Parent}             & \begin{tabular}[c]{@{}l@{}}Knowledge \\ and Events\end{tabular}           & \begin{tabular}[c]{@{}l@{}}Emotions learned at school;\\ Proficiency in describing \\ certain emotions\end{tabular}       & \begin{tabular}[c]{@{}l@{}}C2: \textit{Complex emotions, such as fatigue, are typically} \\ \textit{guided appropriately by kindergarten teachers in class,} \\  \textit{so they should be distinguishable.}\end{tabular}                                                                                                        \\ \cline{2-4} 
                                    & \begin{tabular}[c]{@{}l@{}}Contextual \\ Responses\end{tabular}           & \begin{tabular}[c]{@{}l@{}}Emotional reactions or\\ regulation strategies for\\ different scenarios\end{tabular}          & \begin{tabular}[c]{@{}l@{}}C11: \textit{When he fails in a game, it tends to trigger his} \\ \textit{specific feelings of frustration.}\end{tabular}                                                                                                                                                                   \\ \cline{2-4} 
                                    & \begin{tabular}[c]{@{}l@{}}Behavioral \\ Habits\end{tabular}              & \begin{tabular}[c]{@{}l@{}}Habits of mood adjustment; \\ Seeking external help; \\ Tendency to go to parents\end{tabular} & \begin{tabular}[c]{@{}l@{}}C6: \textit{At present, she cannot calmly face negative} \\ \textit{emotions; she still seeks help from his mother or} \\ \textit{those she feels close to.}\end{tabular}                                                                                                                            \\ \cline{2-4} 
                                    & \begin{tabular}[c]{@{}l@{}}Growth \\ Trends\end{tabular}                  & \begin{tabular}[c]{@{}l@{}}Developmental changes\\ from earlier ages\\\end{tabular}                                                 & \begin{tabular}[c]{@{}l@{}}C12: \textit{When he is happy, he directly expresses his joy.} \\ \textit{When he was younger, he would jump around and} \\ \textit{even scream and shout.}\end{tabular}                                                                                                                             \\ \hline
\multirow{4}{*}{PACEE}                 & \begin{tabular}[c]{@{}l@{}}Specific \\ Emotional \\ Capacity\end{tabular} & \begin{tabular}[c]{@{}l@{}}Understanding, expressing, \\ or positively regulating\\  specific emotions\end{tabular}       & \begin{tabular}[c]{@{}l@{}}C15: \textit{When facing possible failure or competition, she} \\ \textit{considers responding to challenges through effort and} \\ \textit{learning, showing her progress in stressful situations.}\end{tabular}                                                                                    \\ \cline{2-4} 
                                    & \begin{tabular}[c]{@{}l@{}}Macro\\ Emotional \\ Capacity\end{tabular}     & \begin{tabular}[c]{@{}l@{}}Ability to identify emotion\\ origins or others' emotions\end{tabular}                         & \begin{tabular}[c]{@{}l@{}}C2: \textit{...and can recognize the emergence of these} \\ \textit{emotions in specific situations (such as speaking} \\ \textit{on stage). He can feel the impact of the audience's} \\ \textit{reactions on himself. This indicates he can identify} \\ \textit{the source and influence of emotions.}\end{tabular} \\ \cline{2-4} 
                                    & \begin{tabular}[c]{@{}l@{}}Capacity \\ Development \\ Trends\end{tabular} & \begin{tabular}[c]{@{}l@{}}Emotional capacities \\ being developed\end{tabular}                                           & \begin{tabular}[c]{@{}l@{}}C14: \textit{He demonstrates a certain awareness of strategies} \\ \textit{for emotional regulation, indicating he is developing} \\ \textit{the ability to regulate emotions and can attempt to}\\ \textit{use different methods to cope with emotions.}\end{tabular}                                        \\ \cline{2-4} 
                                    & \begin{tabular}[c]{@{}l@{}}Languaguistic \\ Characteristics\end{tabular}  & \begin{tabular}[c]{@{}l@{}}Directness of language \\ expression; Use of non-verbal \\ communication methods\end{tabular}  & \begin{tabular}[c]{@{}l@{}}C13: \textit{He can directly express his emotional state but} \\ \textit{may lack a more extensive and nuanced vocabulary to} \\ \textit{describe his feelings.}\end{tabular}                                                                                                                        \\ \bottomrule[1pt]
\end{tabular}
}
\label{tab:parent_ai_profile_factors}
\end{table}

\subsubsection{PACEE Provides New Insights on Children's Profile for Parental Reflection}
Parents reported that PACEE offered more objective and structured interpretations of children's emotional profiles, revealing perspectives they had previously overlooked. Some realized their earlier judgments were either underestimated or overly optimistic. For example, P2 noted that the AI revealed emotional understanding he had not recognized in his child, while P4 considered the AI's evaluation more rational than her own judgment.

PACEE also helped parents reflect on causes of children's emotional reactions. As P8 explained, ``\textit{He was very reluctant to go on stage. I assumed he was nervous...but the AI suggested that the anxiety might have stemmed from the absence of his mother's support.}''

Overall, most parents perceived these discrepancies between parents and AI as valuable reflections. 14 of 16 parents found PACEE's analysis to be a useful external perspective that complemented their experience. However, two parents questioned the reliability of some interpretations, noting that AI might misread subtle emotions or fail to detect when children are not telling the truth.

\subsubsection{Parents Appreciate Open and Positive Advice from PACEE}
During the study, parents received a total of 513 real-time advice updates from PACEE. Log and video analysis showed that parents most frequently adopted suggestions related to collaborative problem-solving and open-ended questioning (\autoref{appendix:Additional User Evaluation Details}). 

Parents also mentioned the usefulness of open-ended questioning in the interview, as P2 stated, ``\textit{Real-time advice helps me realize some issues I hadn't thought about, such as prompting my child to provide more details. Sometimes I think it's okay when a child says one word, but the advice reminds me to ask for more details.}'' 

Other types of advice, including scenario simulation, concrete suggestions, positive encouragement, and empathy and acceptance, were also adopted by some parents.

\subsubsection{Parents Accept PACEE in Collaborative Guidance, but with Concerns}
Parents generally accepted PACEE as a collaborative assistant in parent–child emotional guidance, while emphasizing that parents should remain the primary decision-makers. Most participants reported that they contributed more to the guidance process, with PACEE mainly providing professional suggestions that parents could selectively adopt.

All parents welcomed PACEE's involvement, though preferred interaction modes varied. Some envisioned PACEE interacting directly with children as a peer-like educator, while others preferred it to support parents only or act as a neutral mediator.

However, parents emphasized clear boundaries for AI involvement. Some rejected features that impose goals on children (P4: ``\textit{Setting goals for children... I think it is up to parents to do it.}''), or intervene in sensitive family matters (P1: ``\textit{Don't press further... for example, on topics that parents find taboo and children find sensitive.}''). Parents also stressed that PACEE should not replace emotional intimacy in parent–child relationships.
% , or require strict compliance with AI advice (P11: ``\textit{If it tells me I have to do something, I definitely won't accept it.}'')

\subsection{RQ3. How Do Parents Perceive PACEE in Addressing Their Challenges in Family Emotion Education?}
\begin{figure}[htb] 
\centering
\includegraphics[width=\columnwidth]{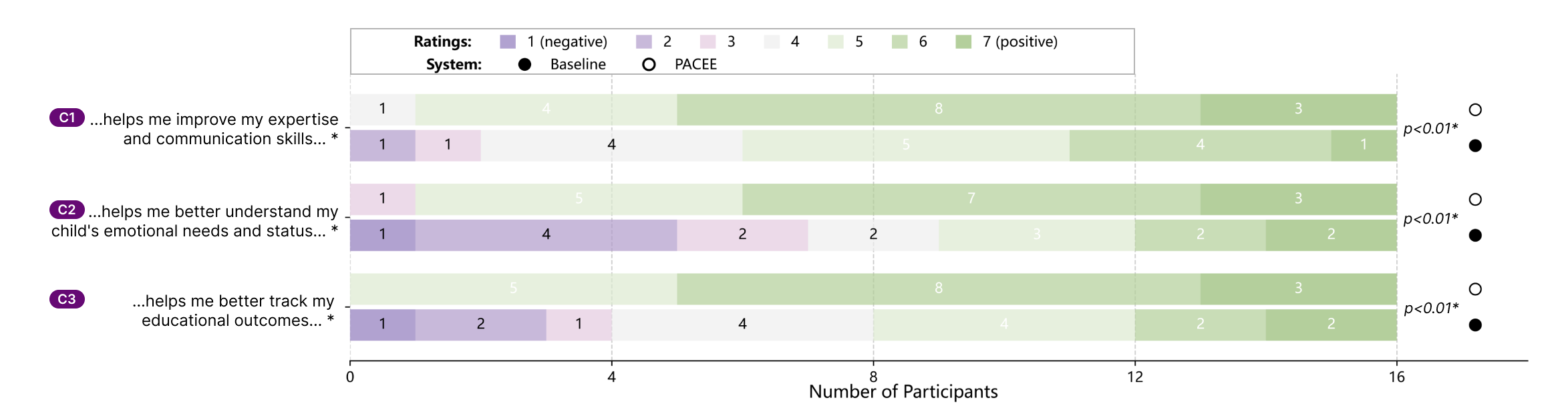}
\caption{Post-questionnaire results on how PACEE and Baseline address the three key challenges.}
\label{figure:questionnaire_data_challenges}
\end{figure}

Parents' ratings show that PACEE significantly outperformed the Baseline in addressing the three key challenges of family emotion education (each $p < 0.01$; \autoref{figure:questionnaire_data_challenges}). Interviews further indicate that parents perceived PACEE as playing multiple supportive roles in mitigating these challenges.

\subsubsection{Enhancing Parental Expertise and Communication Confidence}
Regarding \mybox[defaultpurple]{$C_1$} limited parental expertise, the \textit{Parent-Child Conversational Scenario} module supported parents who reported not knowing ``what to say'' or ``how to guide'' children. Parents experienced difficulties across several steps: recalling experiences ($n=4$), labeling emotions ($n=8$), expressing empathy ($n=8$), reflecting on emotions ($n=5$), and finding positive solutions ($n=6$). 

PACEE's suggestions provided structured guidance across these stages. For example, P8 noted that the step-by-step process enabled smoother interactions, while others reported that the suggestions helped them adopt the child's perspective and steer conversations.

The module also increased parents' confidence: 10 participants reported that the suggestions offered concrete, child-friendly ways to express emotions, helping them organize thoughts and communicate more openly. Most parents reported greater initiative and comfort in emotional conversations. However, some parents noted that overreliance on AI-generated expressions might weaken their own communication abilities.

\subsubsection{Providing Trackable Feedback and Actionable Future Strategies}
Addressing \mybox[defaultpurple]{$C_2$} insufficient feedback, the \textit{Parental Feedback Report} module mitigated issues such as ``lack of timely feedback'' and ``invisible outcomes''. Eight parents reported that AI-generated analyses revealed previously unnoticed shortcomings, prompting self-reflection (e.g., P6: ``\textit{The system reminded me to avoid interrupting. I hadn't realized I was doing that.}''). Five parents also valued structured summaries of conversation highlights, which made parenting efforts more visible and trackable.

Beyond reviewing past interactions, the module provided actionable strategies for future communication. Parents appreciated the evidence-based suggestions and practical perspectives for handling children's emotions. However, some concerns remained, including the limited ability of a single session to reveal long-term outcomes and the need for more personalized suggestions.

\subsubsection{Improving Parental Understanding of Children's Developmental Trajectories}

Regarding \mybox[defaultpurple]{$C_3$}, a lack of understanding of children's emotional states, the \textit{Child Emotional Modeling} module provided data-driven insights into children's emotional traits and behavioral patterns. Twelve parents reported that the system introduced ``new knowledge and perspectives,'' helping them overcome cognitive blind spots and reflect more objectively on their children's emotions (e.g., P12: ``\textit{The system's questions prompted me to think about aspects I hadn't paid attention to before. The conclusions helped me understand my child more systematically and comprehensively.}'' )

Parents also discussed the alignment between their views and the system's analyses: some reported high consistency that strengthened trust, while others found discrepancies useful for reconsidering their perceptions. Several parents further valued that the system contextualized their child's behavior within broader developmental patterns, supporting understanding of age-related characteristics and future development.
\section{DISCUSSION}
In this work, we introduced PACEE, an AI assistant to support parents in family emotion education. We found that PACEE can support their reflection and improve their satisfaction when guiding children's emotions. We clarify that our study does not claim that PACEE improves parents' emotional coaching skills. Instead, it structures parent-child conversations following established emotional coaching steps, helping parents remain engaged and supporting their reflection. Therefore, the improved conversational engagement should not be interpreted as evidence of validated emotional coaching effectiveness, but as a socio-technical system to guide parent-child conversation. Assessing parents' emotional coaching skills would require expert validation, which is beyond the scope of this study and could be pursued in future work.

Here, we further discuss the impact of PACEE on Gottman's Five Steps of Emotion Coaching \cite{gable1999heart}. We then propose broader design implications for parent-centered, AI-supported education systems.

\subsection{Parent-AI Co-construction of Children's Emotional Understanding}
A central challenge in parent-centered AI systems is helping parents better understand their children. Our findings suggest that this understanding should emerge through parent-AI collaboration rather than solely through AI-driven assessment.

Prior systems typically rely either on parent-reported information \cite{zhang2022storybuddy, fang2024edulive, chen2025characterizing} or AI-based assessment \cite{lee2024open}. In contrast, PACEE integrates both: parents contribute experiential knowledge from everyday observations, while AI analyzes parent–child interaction data to identify emotional and developmental characteristics. Therefore, in family contexts, AI may be most effective not as an authority diagnosing children, but as a reflective partner that augments parents' understanding. Future parent-centered AI systems should therefore support the co-construction of child models with parents, combining experiential knowledge with data-driven analysis.

\subsection{Rethinking Gottman's Five Steps as a Triadic AI-Parent-Child Collaboration}
% Our findings suggest that Gottman's Five Steps of Emotion Coaching \cite{gable1999heart} should be reframed, rather than directly implemented, in AI-supported family emotion education for children aged 3-6. 
% Although prior systems have applied this framework in AI-supported contexts \cite{seo2024chacha, ibrahim2025designing, tang2024emoeden}, Gottman's original model assumes a dyadic parent-child interaction in which children can recall, articulate, and reflect on emotional experiences. For preschool children, however, these capacities are still emerging, and parents often face practical challenges in sustaining developmentally appropriate guidance at home. Together, our findings point to a triadic \textit{AI-parent-child} collaboration in which AI supports parents in guiding children's emotional development.

Gottman's model assumes a dyadic parent-child interaction in which children can recall, articulate, and reflect on emotions \cite{gable1999heart} and has been applied in AI-supported HCI contexts, especially in AI-supported emotional education with schoolchildren in a dyadic format \cite{seo2024chacha, ibrahim2025designing, tang2024emoeden}. However, this dyad might not fit well for preschool children aged 3-6 who are still developing their emotional responses to challenges, and their parents also face challenges in sustaining developmentally appropriate guidance at home. Thus, we adapted Gottman's Five Steps of Emotion Coaching and propose a triadic \textit{AI-parent-child} collaboration mode in which AI collaborates with parents to guide their children's emotional development.

\textbf{AI can help parents support emotion recognition by externalizing children's vague or fragmented experiences.} Because preschoolers often struggle to recall and verbalize the causes of emotions \cite{channell2013individual, liwag1995children}, AI-generated scenarios can make emotional situations more concrete. For young children, emotion recognition should be understood less as eliciting verbal explanations and more as helping parents and children jointly interpret emotions through concrete scenarios.

\textbf{Empathy in AI-supported family education can be understood as a shared interpretive process rather than solely an individual parental skill.} Parents often find empathetic listening difficult to sustain in daily interactions. PACEE's \textit{Child Emotional Modeling} module provides an additional perspective on children's emotional states, complementing parents' contextual knowledge with structured developmental cues. This can reduce subjective blind spots and facilitate deeper emotional understanding between parents and children. More broadly, our findings extend empathy from a purely interpersonal skill to a shared interpretive process in which AI supports parents' perspective-taking while parents still lead the emotional relationship.

\textbf{Support for young children's problem solving should prioritize positive-resource scaffolding over solution prescription.} Prior work shows that children aged 3-6 respond well to positive reinforcement \cite{hardy2020using, sigler2005positive}. Building on this insight, PACEE foregrounds positive emotions and strengths through the PERMA model \cite{seligman2010flourish}. In early childhood contexts, the goal of problem-solving should not be narrowly defined as correcting negative behavior or prescribing fixed solutions, but rather as helping children gradually build emotional resources for future coping.

\subsection{Opportunities and Challenges of AI-supported Family Education}
Prior UIST work on AI-supported family and educational systems has largely focused on enabling direct child–AI interaction, positioning children as independent users \cite{chen2025scenic, guo2023sparkybot, cai2023starrypia, zhang2025empowering, li2021faceme, zhao2023narratron, shi2023operar}. A smaller line of work introduces AI into parent-child interaction \cite{vargas2023talemate}, but still assumes that children actively engage with AI, resulting in predominantly symmetric triadic interaction.

In contrast, we introduce an \textit{asymmetric triadic interaction framework}, where AI primarily supports parents while preserving parent-led parent–child interaction. This reframes AI from child-facing to parent-augmenting, reflecting the constraints of sensitive contexts---especially for young children---where direct AI-child interaction may be developmentally inappropriate or risky.

This asymmetric framework can be extended to settings where children face challenges in directly interacting with AI and rely on caregivers as intermediaries. It extends prior work on supporting children with autism spectrum disorder \cite{higuchi2018visualizing, li2025you, washington2016wearable}, cognitive impairments \cite{leonidis2017home}, and language delays \cite{song2016talklime, sohail2025rehnuma}. It can also be applied to sensitive educational situations such as moral guidance, safety-critical contexts, and scenarios involving family privacy.

More broadly, our findings suggest the need to rethink multi-party human–AI interaction beyond symmetric role assumptions. In high-stakes family and educational contexts, rather than engaging all participants equally, \textit{who AI interacts with---and to what extent---} becomes a critical design decision. Designing for asymmetric participation, where AI selectively supports certain roles while remaining minimally intrusive to others, may be essential for ensuring safety and appropriateness.

At the same time, applying generative AI in such sensitive contexts raises important challenges. Because PACEE's guidance is AI-generated and lacks expert validation, incorrect or poorly contextualized suggestions may interfere with effective emotional coaching. These risks highlight the importance of positioning systems like PACEE as assistive tools that augment caregivers' reflection rather than replace professional expertise.

Taken together, PACEE serves as a design probe exploring how generative AI may support parent-led education in everyday family contexts. Future systems should integrate insights from developmental psychology and incorporate expert-informed safeguards.

\subsection{Design Considerations for Parent-centered, AI-supported Education Systems}
Parent interviews and prior work suggest three design considerations for \textit{parent-centered, AI-supported education systems}.

\textbf{Customize AI support for parents as well as children.}
While prior work primarily adapts AI coaching to children's profiles \cite{clabaugh2019long, tang2024emoeden, zhang2022storybuddy}, our findings show that parents also vary in their expectations regarding the difficulty, timing, and frequency of AI assistance. Future systems should therefore adapt AI support to both children's characteristics and parental preferences.

\textbf{Enable transparent and negotiable parent-AI collaboration.}
Parents expressed concerns about the reliability of AI interpretations and potential disagreement with AI assessments. Prior work suggests that transparency improves human–AI collaboration \cite{liao2020questioning}. Systems should therefore explain how suggestions are generated and allow parents to question or adjust AI interpretations.

\textbf{Respect parental boundaries and preserve parental autonomy.}
Parents consistently viewed themselves as primary decision-makers in their children's education and were cautious about AI overstepping family boundaries. Consistent with prior work on parental decision-making \cite{sisk2020parental}, our findings extend this insight to emotion-sensitive contexts in early childhood education. AI systems should avoid overly prescriptive guidance and instead support flexible, parent-controlled use.

\subsection{Limitations and Future Work}
This study has several limitations. First, participants were limited to Chinese families, and most parent participants were mothers, reflecting common caregiving patterns for children aged 3–6 \cite{mcbride1993comparison, schoppe2013comparisons, hernawati2020differences}. Future work should examine broader cultural contexts and include more balanced parental roles. Second, our design cannot fully isolate the foundation model variable. While our focus is the interaction framework rather than raw model output (\autoref{section: Baseline Condition}), future work could adopt a $2\times2$ factorial design--(Interface: Baseline vs.\ PACEE) $\times$ (Model: Doubao vs.\ GPT-4o)--to separate interface and model effects. Finally, the study was conducted in a lab setting using a touchscreen laptop. Future work should conduct longitudinal in-home deployments to examine parent–AI collaboration in everyday family environments.
\section{CONCLUSION}
We presented PACEE, an LLM-based assistant that helps parents guide their children's emotional development. In a study with 16 parent–child pairs, PACEE enhanced parent–child engagement, fostering deeper emotional exchanges while alleviating parental challenges. We hope this work advances emotion coaching research and inspires the design of parent-centered, AI-supported family education systems.

% \input{data/chap10}
%TC:ignore
\bibliographystyle{ACM-Reference-Format}
\bibliography{ref/main_citation}

@article{ku2007mining,
  title={Mining opinions from the Web: Beyond relevance retrieval},
  author={Ku, Lun-Wei and Chen, Hsin-Hsi},
  journal={Journal of the American Society for Information Science and Technology},
  volume={58},
  number={12},
  pages={1838--1850},
  year={2007},
  publisher={Wiley Online Library}
}

@inproceedings{tang2024emoeden,
  title={Emoeden: Applying generative artificial intelligence to emotional learning for children with high-function autism},
  author={Tang, Yilin and Chen, Liuqing and Chen, Ziyu and Chen, Wenkai and Cai, Yu and Du, Yao and Yang, Fan and Sun, Lingyun},
  booktitle={Proceedings of the 2024 CHI Conference on Human Factors in Computing Systems},
  pages={1--20},
  year={2024}
}

@article{sun2024exploring,
  title={Exploring Parent's Needs for Children-Centered AI to Support Preschoolers' Storytelling and Reading Activities},
  author={Sun, Yuling and Liu, Jiali and Yao, Bingsheng and Chen, Jiaju and Wang, Dakuo and Ma, Xiaojuan and Lu, Yuxuan and Xu, Ying and He, Liang},
  journal={arXiv e-prints},
  pages={arXiv--2401},
  year={2024}
}

@article{shen2025easel,
  title={eaSEL: Promoting Social-Emotional Learning and Parent-Child Interaction through AI-Mediated Content Consumption},
  author={Shen, Jocelyn and Chen, Jennifer King and Findlater, Leah and Smith, Griffin Dietz},
  journal={arXiv preprint arXiv:2501.17819},
  year={2025}
}

@article{choi2024aacesstalk,
  title={AACessTalk: Fostering Communication between Minimally Verbal Autistic Children and Parents with Contextual Guidance and Card Recommendation},
  author={Choi, Dasom and Park, SoHyun and Lee, Kyungah and Hong, Hwajung and Kim, Young-Ho},
  journal={arXiv preprint arXiv:2409.09641},
  year={2024}
}

@inproceedings{santos2020therapist,
  title={Therapist vibe: children's expressions of their emotions through storytelling with a chatbot},
  author={Santos, Kyle-Althea and Ong, Ethel and Resurreccion, Ron},
  booktitle={Proceedings of the interaction design and children conference},
  pages={483--494},
  year={2020}
}

@article{rothbart2004temperament,
  title={Temperament and self-regulation},
  author={Rothbart, Mary K and Ellis, Lesa K and Posner, Michael I},
  journal={Handbook of self-regulation: Research, theory, and applications},
  volume={2},
  pages={441--460},
  year={2004}
}

@book{dewey2024democracy,
  title={Democracy and education},
  author={Dewey, John},
  year={2024},
  publisher={Columbia University Press}
}

@article{pons2004emotion,
  title={Emotion comprehension between 3 and 11 years: Developmental periods and hierarchical organization},
  author={Pons, Francisco and Harris, Paul L and De Rosnay, Marc},
  journal={European journal of developmental psychology},
  volume={1},
  number={2},
  pages={127--152},
  year={2004},
  publisher={Taylor \& Francis}
}

@article{albanese2006children,
  title={Children’s emotion understanding: preliminary data from the Italian validation project of Test of Emotion Comprehension (TEC)},
  author={Albanese, Ottavia and Grazzani, Ilaria and Molina, Paola and Antoniotti, Carla and Arati, Laura and Farina, Eleonora and Pons, Francisco and others},
  journal={Toward emotional competences},
  pages={39--53},
  year={2006},
  publisher={Aalborg University Press Aalborg, Denmark}
}

@article{denham1997parental,
  title={Parental contributions to preschoolers' emotional competence: Direct and indirect effects},
  author={Denham, Susanne A and Mitchell-Copeland, Jennifer and Strandberg, Katherine and Auerbach, Sharon and Blair, Kimberly},
  journal={Motivation and emotion},
  volume={21},
  pages={65--86},
  year={1997},
  publisher={Springer}
}

@article{mavroveli2008investigation,
  title={Investigation of the construct of trait emotional intelligence in children},
  author={Mavroveli, Stella and Petrides, Konstantinos V and Shove, Chloe and Whitehead, Amanda},
  journal={European child \& adolescent psychiatry},
  volume={17},
  number={8},
  pages={516--526},
  year={2008},
  publisher={Springer}
}

@article{nicolaidou2022gamified,
  title={A gamified app on emotion recognition and anger management for pre-school children},
  author={Nicolaidou, Iolie and Tozzi, Federica and Antoniades, Athos},
  journal={International Journal of Child-Computer Interaction},
  volume={31},
  pages={100449},
  year={2022},
  publisher={Elsevier}
}

@article{bullock1985further,
  title={Further evidence on preschoolers' interpretation of facial expressions},
  author={Bullock, Merry and Russell, James A},
  journal={International Journal of Behavioral Development},
  volume={8},
  number={1},
  pages={15--38},
  year={1985},
  publisher={Sage Publications Sage CA: Thousand Oaks, CA}
}

@article{barden1980children,
  title={Children's consensual knowledge about the experiential determinants of emotion.},
  author={Barden, R Christopher and Zelko, Frank A and Duncan, S Wayne and Masters, John C},
  journal={Journal of personality and Social Psychology},
  volume={39},
  number={5},
  pages={968},
  year={1980},
  publisher={American Psychological Association}
}

@article{harris1989young,
  title={Young children's theory of mind and emotion},
  author={Harris, Paul L and Johnson, Carl N and Hutton, Deborah and Andrews, Giles and Cooke, Tim},
  journal={Cognition \& Emotion},
  volume={3},
  number={4},
  pages={379--400},
  year={1989},
  publisher={Taylor \& Francis}
}

@article{fonagy1997relationship,
  title={The relationship between belief-desire reasoning and a projective measure of attachment security (SAT)},
  author={Fonagy, Peter and Redfern, Sheila and Charman, Tony},
  journal={British Journal of Developmental Psychology},
  volume={15},
  number={1},
  pages={51--61},
  year={1997},
  publisher={Wiley Online Library}
}

@article{harris1983children,
  title={Children's understanding of the link between situation and emotion},
  author={Harris, Paul L},
  journal={Journal of experimental child psychology},
  volume={36},
  number={3},
  pages={490--509},
  year={1983},
  publisher={Elsevier}
}

@article{altshuler1989developmental,
  title={Developmental changes in children's awareness of strategies for coping with uncontrollable stress},
  author={Altshuler, Jennifer L and Ruble, Diane N},
  journal={Child development},
  pages={1337--1349},
  year={1989},
  publisher={JSTOR}
}

@article{gardner1988japanese,
  title={Japanese children's understanding of the distinction between real and apparent emotion},
  author={Gardner, D and Harris, PL and Ohmoto, Mi and Hamazaki, T},
  journal={International Journal of Behavioral Development},
  volume={11},
  number={2},
  pages={203--218},
  year={1988},
  publisher={Sage Publications Sage CA: Thousand Oaks, CA}
}

@article{arsenio1995children,
  title={Children's conceptions of sociomoral affect: Happy victimizers, mixed emotions, and other expectancies.},
  author={Arsenio, William and Lover, Anthony},
  year={1995},
  publisher={Cambridge University Press}
}

@article{harter1989developmental,
  title={Developmental changes in children’s understanding of single, multiple, and blended emotion concepts},
  author={Harter, Susan and Whitesell, Nancy Rumbaugh},
  journal={Children’s understanding of emotion},
  pages={81--116},
  year={1989}
}

@inproceedings{seo2024chacha,
  title={Chacha: leveraging large language models to prompt children to share their emotions about personal events},
  author={Seo, Woosuk and Yang, Chanmo and Kim, Young-Ho},
  booktitle={Proceedings of the 2024 CHI Conference on Human Factors in Computing Systems},
  pages={1--20},
  year={2024}
}

@book{gable1999heart,
  title={The Heart of Parenting: How to Raise an Emotionally Intelligent Child},
  author={Gottman, J.M. and DeClaire, J.},
  isbn={9780684801308},
  lccn={96038947},
  series={Simon \& Schuster paperbacks},
  url={https://books.google.co.jp/books?id=64pdPgAACAAJ},
  year={1997},
  publisher={Simon \& Schuster}
}

@article{saarni1999development,
  title={The development of emotional competence Guilford Press},
  author={Saarni, C},
  journal={New York, NY},
  year={1999}
}

@article{alegre2011parenting,
  title={Parenting styles and children’s emotional intelligence: What do we know?},
  author={Alegre, Alberto},
  journal={The Family Journal},
  volume={19},
  number={1},
  pages={56--62},
  year={2011},
  publisher={Sage Publications Sage CA: Los Angeles, CA}
}

@article{havighurst2021tuning,
  title={Tuning in to Kids: An emotion coaching approach to working with parents},
  author={Havighurst, Sophie S and Kehoe, Christiane E},
  journal={Family-based intervention for child and adolescent mental health: A core competencies approach},
  pages={269--283},
  year={2021},
  publisher={Cambridge University Press}
}

@article{wang2021protection,
  title={Protection or punishment? relating the design space of parental control apps and perceptions about them to support parenting for online safety},
  author={Wang, Ge and Zhao, Jun and Van Kleek, Max and Shadbolt, Nigel},
  journal={Proceedings of the ACM on Human-Computer Interaction},
  volume={5},
  number={CSCW2},
  pages={1--26},
  year={2021},
  publisher={ACM New York, NY, USA}
}

@article{bartlett1982emotional,
  title={Emotional mood and memory in young children},
  author={Bartlett, James C and Burleson, Georgia and Santrock, John W},
  journal={Journal of Experimental Child Psychology},
  volume={34},
  number={1},
  pages={59--76},
  year={1982},
  publisher={Elsevier}
}

@article{wulandari2021development,
  title={THE DEVELOPMENT OF CHILDREN'S EMOTIONAL: A SYSTEMATIC LITERATURE REVIEW},
  author={Wulandari, Tri},
  journal={OPTIMA: Journal Of Guidance and Counseling},
  volume={1},
  number={2},
  pages={13--27},
  year={2021}
}

@article{fadlillah2022analysis,
  title={Analysis of Diana Baumrind's Parenting Style on Early Childhood Development},
  author={Fadlillah, Muhammad and Fauziah, Syifa},
  journal={Al-Ishlah: Jurnal Pendidikan},
  volume={14},
  number={2},
  pages={2127--2134},
  year={2022}
}

@article{cole1994development,
  title={The development of emotion regulation and dysregulation: A clinical perspective},
  author={Cole, Pamela M and Michel, Margaret K and Teti, Laureen O'Donnell},
  journal={Monographs of the society for research in child development},
  pages={73--100},
  year={1994},
  publisher={JSTOR}
}

@book{kahneman2011thinking,
  title={Thinking, fast and slow},
  author={Kahneman, Daniel},
  year={2011},
  publisher={macmillan}
}

@article{christakopoulou2024agents,
  title={Agents thinking fast and slow: A talker-reasoner architecture},
  author={Christakopoulou, Konstantina and Mourad, Shibl and Matari{\'c}, Maja},
  journal={arXiv preprint arXiv:2410.08328},
  year={2024}
}

@article{lin2023swiftsage,
  title={Swiftsage: A generative agent with fast and slow thinking for complex interactive tasks},
  author={Lin, Bill Yuchen and Fu, Yicheng and Yang, Karina and Brahman, Faeze and Huang, Shiyu and Bhagavatula, Chandra and Ammanabrolu, Prithviraj and Choi, Yejin and Ren, Xiang},
  journal={Advances in Neural Information Processing Systems},
  volume={36},
  pages={23813--23825},
  year={2023}
}

@book{gibaud1978development,
  title={Development and utility of the Parenting Sense of Competence Scale},
  author={Gibaud-Wallston, Jonatha and Wandersmann, Lois Pall},
  year={1978},
  publisher={John F. Kennedy center for research on education and human development}
}

@incollection{hart1988development,
  title={Development of NASA-TLX (Task Load Index): Results of empirical and theoretical research},
  author={Hart, Sandra G and Staveland, Lowell E},
  booktitle={Advances in psychology},
  volume={52},
  pages={139--183},
  year={1988},
  publisher={Elsevier}
}

@misc{bandura1997self,
  title={Self-efficacy: The exercise of control},
  author={Bandura, Albert},
  year={1997},
  publisher={Freeman}
}

@article{sandler2011long,
  title={Long-term impact of prevention programs to promote effective parenting: Lasting effects but uncertain processes},
  author={Sandler, Irwin N and Schoenfelder, Erin N and Wolchik, Sharlene A and MacKinnon, David P},
  journal={Annual review of psychology},
  volume={62},
  number={1},
  pages={299--329},
  year={2011},
  publisher={Annual Reviews}
}

@article{castro2015parents,
  title={Parents' emotion-related beliefs, behaviours, and skills predict children's recognition of emotion},
  author={Castro, Vanessa L and Halberstadt, Amy G and Lozada, Fantasy T and Craig, Ashley B},
  journal={Infant and child development},
  volume={24},
  number={1},
  pages={1--22},
  year={2015},
  publisher={Wiley Online Library}
}

@article{sanders1999triple,
  title={Triple P-Positive Parenting Program: Towards an empirically validated multilevel parenting and family support strategy for the prevention of behavior and emotional problems in children},
  author={Sanders, Matthew R},
  journal={Clinical child and family psychology review},
  volume={2},
  pages={71--90},
  year={1999},
  publisher={Springer}
}

@article{clarke2017thematic,
  title={Thematic analysis},
  author={Clarke, Victoria and Braun, Virginia},
  journal={The journal of positive psychology},
  volume={12},
  number={3},
  pages={297--298},
  year={2017},
  publisher={Taylor \& Francis}
}

@article{yu2025suitability,
  title={Suitability of Chinese GenAI Platforms for Early Childhood Education: A Multifaceted Evaluation},
  author={Yu, Sihang and Hou, Yichen and Li, Hui},
  year={2025}
}

@misc{newsDoubao1,  
  author = {Ben Jiang},  
  title = {China’s hottest AI bot? ByteDance’s Doubao tops the charts with 51 million active users},  
  year = {2024},  
  howpublished = {\url{https://www.scmp.com/tech/tech-trends/article/3286276/chinas-hottest-ai-bot-bytedances-doubao-tops-charts-51-million-active-users}},  
  note = {Accessed: 2025-05-01}  
}

@misc{newsDoubao2,  
  author = {Arooj Ahmed},  
  title = {China’s AI Chatbot Market Sees ByteDance’s Doubao Leading Through Innovation and Accessibility},  
  year = {2025},  
  howpublished = {\url{https://www.digitalinformationworld.com/2025/01/chinas-ai-chatbot-market-sees.html}},  
  note = {Accessed: 2025-05-01}  
}

@article{clabaugh2019long,
  title={Long-term personalization of an in-home socially assistive robot for children with autism spectrum disorders},
  author={Clabaugh, Caitlyn and Mahajan, Kartik and Jain, Shomik and Pakkar, Roxanna and Becerra, David and Shi, Zhonghao and Deng, Eric and Lee, Rhianna and Ragusa, Gisele and Matari{\'c}, Maja},
  journal={Frontiers in Robotics and AI},
  volume={6},
  pages={110},
  year={2019},
  publisher={Frontiers Media SA}
}

@inproceedings{liao2020questioning,
  title={Questioning the AI: informing design practices for explainable AI user experiences},
  author={Liao, Q Vera and Gruen, Daniel and Miller, Sarah},
  booktitle={Proceedings of the 2020 CHI conference on human factors in computing systems},
  pages={1--15},
  year={2020}
}

@article{sisk2020parental,
  title={Parental attitudes toward artificial intelligence-driven precision medicine technologies in pediatric healthcare},
  author={Sisk, Bryan A and Antes, Alison L and Burrous, Sara and DuBois, James M},
  journal={Children},
  volume={7},
  number={9},
  pages={145},
  year={2020},
  publisher={MDPI}
}

@article{fu2022social,
  title={Social Emotional Learning with Conversational Agents: Reviewing Current Designs and Probing Parents' Ideas for Future Ones},
  author={Fu, Yue and Michelson, Rebecca and Lin, Yifan and Nguyen, Lynn K and Tayebi, Tala June and Hiniker, Alexis},
  journal={Proceedings of the ACM on Interactive, Mobile, Wearable and Ubiquitous Technologies},
  volume={6},
  number={2},
  pages={1--23},
  year={2022},
  publisher={ACM New York, NY, USA}
}

@article{baker2017positive,
  title={Positive early childhood education: Expanding the reach of positive psychology into early childhood},
  author={Baker, Lisa and Green, Suzy and Falecki, Daniela},
  journal={European Journal of Applied Positive Psychology},
  volume={1},
  number={8},
  pages={1--12},
  year={2017}
}

@inproceedings{fan2018emostory,
  title={Emostory: A game-based system supporting children's emotional development},
  author={Fan, Min and Fan, Jianyu and Jin, Sheng and Antle, Alissa N and Pasquier, Philippe},
  booktitle={Extended Abstracts of the 2018 CHI conference on human factors in computing systems},
  pages={1--6},
  year={2018}
}

@inproceedings{kim2018can,
  title={Can a machine tend to teenagers' emotional needs? A study with conversational agents},
  author={Kim, Junhan and Kim, Yoojung and Kim, Byungjoon and Yun, Sukyung and Kim, Minjoon and Lee, Joongseek},
  booktitle={Extended abstracts of the 2018 CHI conference on human factors in computing systems},
  pages={1--6},
  year={2018}
}

@inproceedings{ahmadpour2023understanding,
  title={Understanding how technology can support social-emotional learning of children: a dyadic trauma-informed participatory design with proxies},
  author={Ahmadpour, Naseem and Loke, Lian and Gray, Carl and Cao, Yidan and Macdonald, Chloe and Hart, Rebecca},
  booktitle={Proceedings of the 2023 CHI Conference on Human Factors in Computing Systems},
  pages={1--17},
  year={2023}
}

@article{gottman1996parental,
  title={Parental meta-emotion philosophy and the emotional life of families: theoretical models and preliminary data.},
  author={Gottman, John M and Katz, Lynn Fainsilber and Hooven, Carole},
  journal={Journal of family psychology},
  volume={10},
  number={3},
  pages={243},
  year={1996},
  publisher={American Psychological Association}
}

@article{BROWN2022126,
title = {The sound of self-regulation: Music program relates to an advantage for children at risk},
journal = {Early Childhood Research Quarterly},
volume = {60},
pages = {126-136},
year = {2022},
issn = {0885-2006},
doi = {https://doi.org/10.1016/j.ecresq.2022.01.002},
url = {https://www.sciencedirect.com/science/article/pii/S0885200622000023},
author = {Eleanor D. Brown and Mary Ann Blumenthal and Alyssa A. Allen}
}

@inproceedings{sharma2025robot,
  title={A robot teacher" is very good for learning, but not for human relationships": Japanese Children's Critical Perspectives Towards Ethical AI Futures},
  author={Sharma, Sumita and Klemettil{\"a}, Pauli and Tanaka, Junko},
  booktitle={CHI'25: Proceedings of the 2025 CHI Conference on Human Factors in Computing Systems},
  year={2025},
  organization={ACM}
}

@inproceedings{quan2025parents,
  title={Parents, Children, and ChatGPT in Home Environments: The Conversation Content and the Interaction Mode},
  author={Quan, Shuang and Du, Yao and Lyu, Yao},
  booktitle={Proceedings of the Extended Abstracts of the CHI Conference on Human Factors in Computing Systems},
  pages={1--6},
  year={2025}
}

@article{bernacki2021systematic,
  title={A systematic review of research on personalized learning: Personalized by whom, to what, how, and for what purpose (s)?},
  author={Bernacki, Matthew L and Greene, Meghan J and Lobczowski, Nikki G},
  journal={Educational Psychology Review},
  volume={33},
  number={4},
  pages={1675--1715},
  year={2021},
  publisher={Springer}
}

@article{wellman2004scaling,
  title={Scaling of theory-of-mind tasks},
  author={Wellman, Henry M and Liu, David},
  journal={Child development},
  volume={75},
  number={2},
  pages={523--541},
  year={2004},
  publisher={Wiley Online Library}
}

@article{calvo2010affect,
  title={Affect detection: An interdisciplinary review of models, methods, and their applications},
  author={Calvo, Rafael A and D'Mello, Sidney},
  journal={IEEE Transactions on affective computing},
  volume={1},
  number={1},
  pages={18--37},
  year={2010},
  publisher={IEEE}
}

@article{gottman2005five,
  title={Five Steps of Emotion Coaching},
  author={Gottman, John},
  journal={Seattle, WA: Talaris Research},
  year={2005}
}

@article{kovich2023application,
  title={Application of the PERMA model of well-being in undergraduate students},
  author={Kovich, Melissa K and Simpson, Vicki L and Foli, Karen J and Hass, Zachary and Phillips, Rhonda G},
  journal={International journal of community well-being},
  volume={6},
  number={1},
  pages={1--20},
  year={2023},
  publisher={Springer}
}

@article{nebrida2018m,
  title={I’m perfectly imperfect”: Exploring the relationship between PERMA model of wellbeing with self-esteem among persons with disabilities},
  author={Nebrida, Jayson D and Dullas, Angelo},
  journal={International Journal of Research Studies in Psychology},
  volume={7},
  number={2},
  pages={27--44},
  year={2018}
}

@article{chisale2022perma,
  title={PERMA Model and Mental Health Practice},
  author={Chisale, Enipher and Phiri, FE},
  journal={Asian Journal of Pharmacy Nursing},
  volume={10},
  number={2},
  pages={21--24},
  year={2022}
}

@article{seligman2010flourish,
  title={Flourish: Positive psychology and positive interventions},
  author={Seligman, Martin},
  journal={The Tanner lectures on human values},
  volume={31},
  number={4},
  pages={1--56},
  year={2010},
  publisher={Citeseer}
}

@inproceedings{ibrahim2025designing,
  title={Designing Daily Supports for Parent-Child Conversations about Emotion: Ecological Momentary Assessment as Intervention},
  author={Ibrahim, Seray and Klasnja, Predrag and Gross, James J and Slovak, Petr},
  booktitle={CHI Conference on Human Factors in Computing Systems (CHI’25)},
  pages={1--21},
  year={2025},
  organization={Association for Computing Machinery}
}

@inproceedings{hightower2019exploring,
  title={Exploring parent use of early STEM media to inform design for children},
  author={Hightower, Brianna and Sheehan, Kelly and Lauricella, Alexis and Wartella, Ellen},
  booktitle={Proceedings of the 18th ACM International Conference on Interaction Design and Children},
  pages={102--108},
  year={2019}
}

@article{fang2024edulive,
  title={EduLive: Re-Creating Cues for Instructor-Learners Interaction in Educational Live Streams with Learners' Transcript-Based Annotations},
  author={Fang, Jingchao and Park, Jeongeon and Kim, Juho and Wang, Hao-Chuan},
  journal={Proceedings of the ACM on Human-Computer Interaction},
  volume={8},
  number={CSCW2},
  pages={1--33},
  year={2024},
  publisher={ACM New York, NY, USA}
}

@inproceedings{lee2024open,
  title={Open Sesame? Open Salami! Personalizing Vocabulary Assessment-Intervention for Children via Pervasive Profiling and Bespoke Storybook Generation},
  author={Lee, Jungeun and Yoon, Suwon and Lee, Kyoosik and Jeong, Eunae and Cho, Jae-Eun and Park, Wonjeong and Yim, Dongsun and Hwang, Inseok},
  booktitle={Proceedings of the 2024 CHI Conference on Human Factors in Computing Systems},
  pages={1--32},
  year={2024}
}

@article{chen2025characterizing,
  title={Characterizing LLM-Empowered Personalized Story-Reading and Interaction for Children: Insights from Multi-Stakeholder Perspectives},
  author={Chen, Jiaju and Tang, Minglong and Lu, Yuxuan and Yao, Bingsheng and Fan, Elissa and Ma, Xiaojuan and Xu, Ying and Wang, Dakuo and Sun, Yuling and He, Liang},
  journal={arXiv preprint arXiv:2503.00590},
  year={2025}
}

@inproceedings{zhang2022storybuddy,
  title={Storybuddy: A human-ai collaborative chatbot for parent-child interactive storytelling with flexible parental involvement},
  author={Zhang, Zheng and Xu, Ying and Wang, Yanhao and Yao, Bingsheng and Ritchie, Daniel and Wu, Tongshuang and Yu, Mo and Wang, Dakuo and Li, Toby Jia-Jun},
  booktitle={Proceedings of the 2022 CHI Conference on Human Factors in Computing Systems},
  pages={1--21},
  year={2022}
}

@inproceedings{jonas2023supporting,
  title={Supporting from the Background: How a Mobile Application for Parent Skills Development Encourages Authoritative Parenting},
  author={Jonas, Rebecca M and Hanrahan, Benjamin V and Sloan, Carlie J and Fosco, Gregory M},
  booktitle={Companion Publication of the 2023 Conference on Computer Supported Cooperative Work and Social Computing},
  pages={105--111},
  year={2023}
}

@inproceedings{xiao2012supporting,
  title={Supporting parent-young child activities with interactive tabletops: a conceptual analysis},
  author={Xiao, Lu and Martin, Jennifer},
  booktitle={Proceedings of the ACM 2012 conference on Computer Supported Cooperative Work Companion},
  pages={301--310},
  year={2012}
}

@inproceedings{yu2024parent,
  title={Parent-Child Joint Media Engagement Within HCI: A Scoping Analysis of the Research Landscape},
  author={Yu, Junnan and Qi, Xiang and Yang, Siqi},
  booktitle={Proceedings of the 2024 CHI Conference on Human Factors in Computing Systems},
  pages={1--21},
  year={2024}
}

@inproceedings{xu2023mathkingdom,
  title={Mathkingdom: Teaching children mathematical language through speaking at home via a voice-guided game},
  author={Xu, Wenjie and Ma, Jiayi and Yao, Jiayu and Lin, Weijia and Zhang, Chao and Xia, Xuanhe and Zhuang, Nan and Weng, Shitong and Xie, Xiaoqian and Feng, Shuyue and others},
  booktitle={Proceedings of the 2023 CHI Conference on Human Factors in Computing Systems},
  pages={1--14},
  year={2023}
}

@inproceedings{zhang2024mathemyths,
  title={Mathemyths: leveraging large language models to teach mathematical language through Child-AI co-creative storytelling},
  author={Zhang, Chao and Liu, Xuechen and Ziska, Katherine and Jeon, Soobin and Yu, Chi-Lin and Xu, Ying},
  booktitle={Proceedings of the 2024 CHI Conference on Human Factors in Computing Systems},
  pages={1--23},
  year={2024}
}

@article{toran2024parent,
  title={The parent--child relationship in the digital era: The mediator role of digital parental awareness},
  author={Toran, Mehmet and Kulaks{\i}z, Taibe and {\"O}zden, B{\"u}lent},
  journal={Children and Youth Services Review},
  volume={161},
  pages={107617},
  year={2024},
  publisher={Elsevier}
}

@article{morawska2023managing,
  title={Managing screen use in the under-fives: Recommendations for parenting intervention development},
  author={Morawska, Alina and Mitchell, Amy E and Tooth, Leigh R},
  journal={Clinical Child and Family Psychology Review},
  volume={26},
  number={4},
  pages={943--956},
  year={2023},
  publisher={Springer}
}

@article{corwin2012secure,
  title={A secure attachment base is ideal to be a great learner},
  author={Corwin, Heather L},
  journal={Journal of Prenatal \& Perinatal Psychology \& Health},
  volume={27},
  number={1},
  pages={38},
  year={2012},
  publisher={Association for Pre \& Perinatal Psychology and Health}
}

@article{brock2016interparental,
  title={Interparental conflict, children's security with parents, and long-term risk of internalizing problems: A longitudinal study from ages 2 to 10},
  author={Brock, Rebecca L and Kochanska, Grazyna},
  journal={Development and psychopathology},
  volume={28},
  number={1},
  pages={45--54},
  year={2016},
  publisher={Cambridge University Press}
}

@article{meng2020effects,
  title={Effects of parental empathy and emotion regulation on social competence and emotional/behavioral problems of school-age children},
  author={Meng, Kun and Yuan, Yizhe and Wang, Yali and Liang, Jianning and Wang, Lijun and Shen, Jianfei and Wang, Yanyu},
  journal={Pediatric investigation},
  volume={4},
  number={02},
  pages={91--98},
  year={2020},
  publisher={Chinese Medical Journals Publishing House Co., Ltd. 42 Dongsi Xidajie~…}
}

@inproceedings{grassl2024coding,
  title={Coding to Cope: Teaching Programming to Children with Emotional and Behavioral Disorders},
  author={Gra{\ss}l, Isabella and Fraser, Gordon},
  booktitle={Proceedings of the 46th International Conference on Software Engineering: Software Engineering Education and Training},
  pages={127--138},
  year={2024}
}

@inproceedings{slovak2016scaffolding,
  title={Scaffolding the scaffolding: supporting children's social-emotional learning at home},
  author={Slov{\'a}k, Petr and Rowan, Kael and Frauenberger, Christopher and Gilad-Bachrach, Ran and Doces, Mia and Smith, Brian and Kamb, Rachel and Fitzpatrick, Geraldine},
  booktitle={Proceedings of the 19th ACM conference on computer-supported cooperative work \& social computing},
  pages={1751--1765},
  year={2016}
}

@inproceedings{dumaru2023after,
  title={“After she fell asleep, it went to my next podcast, which was about a serial killer”: Unveiling Needs and Expectations Regarding Parental Control within Digital Assistant},
  author={Dumaru, Prakriti and Al-Ameen, Mahdi Nasrullah},
  booktitle={Companion Publication of the 2023 Conference on Computer Supported Cooperative Work and Social Computing},
  pages={17--21},
  year={2023}
}

@inproceedings{cingel2017parents,
  title={How parents engage children in tablet-based reading experiences: An exploration of haptic feedback},
  author={Cingel, Drew and Piper, Anne Marie},
  booktitle={Proceedings of the 2017 ACM Conference on Computer Supported Cooperative Work and Social Computing},
  pages={505--510},
  year={2017}
}

@inproceedings{nikkhah2021helping,
  title={Helping Their Child, Helping Each Other: Parents’ Mediated Social Support in the Children's Hospital},
  author={Nikkhah, Sarah and John, Swaroop and Yalamarti, Krishna Supradeep and L. Mueller, Emily and D. Miller, Andrew},
  booktitle={Companion Publication of the 2021 Conference on Computer Supported Cooperative Work and Social Computing},
  pages={140--143},
  year={2021}
}

@inproceedings{hiniker2016not,
  title={Not at the dinner table: Parents' and children's perspectives on family technology rules},
  author={Hiniker, Alexis and Schoenebeck, Sarita Y and Kientz, Julie A},
  booktitle={Proceedings of the 19th ACM conference on computer-supported cooperative work \& social computing},
  pages={1376--1389},
  year={2016}
}

@inproceedings{mazmanian2017okay,
  title={" Okay, One More Episode" An Ethnography of Parenting in the Digital Age},
  author={Mazmanian, Melissa and Lanette, Simone},
  booktitle={Proceedings of the 2017 ACM conference on computer supported cooperative work and social computing},
  pages={2273--2286},
  year={2017}
}

@inproceedings{chen2020new,
  title={A new approach to parallel interaction through co-located and object-oriented storytelling},
  author={Chen, Bo-Han and Wong, Sai-Keung and Chang, Wei-Che},
  booktitle={Companion Publication of the 2020 Conference on Computer Supported Cooperative Work and Social Computing},
  pages={233--238},
  year={2020}
}

@inproceedings{ghiotti2023prototyping,
  title={Prototyping Kodi: Defining Design Requirements to Develop a Virtual Chat-bot for Autistic Children and Their Caregivers},
  author={Ghiotti, Narayan and Clulow, David and Cheon, Serene and Cui, Kevin and Kang, Hyo},
  booktitle={Companion Publication of the 2023 Conference on Computer Supported Cooperative Work and Social Computing},
  pages={126--131},
  year={2023}
}

@inproceedings{sharma2016promoting,
  title={Promoting joint attention with computer supported collaboration in children with autism},
  author={Sharma, Sumita and Srivastava, Saurabh and Achary, Krishnaveni and Varkey, Blessin and Heimonen, Tomi and Hakulinen, Jaakko Samuli and Turunen, Markku and Rajput, Nitendra},
  booktitle={Proceedings of the 19th ACM conference on computer-supported cooperative work \& social computing},
  pages={1560--1571},
  year={2016}
}

@inproceedings{yarosh2013almost,
  title={" almost touching" parent-child remote communication using the sharetable system},
  author={Yarosh, Svetlana and Tang, Anthony and Mokashi, Sanika and Abowd, Gregory D},
  booktitle={Proceedings of the 2013 conference on Computer supported cooperative work},
  pages={181--192},
  year={2013}
}

@article{mcbride1993comparison,
  title={A comparison of mother and father involvement with their preschool age children},
  author={McBride, Brent A and Mills, Gail},
  journal={Early childhood research quarterly},
  volume={8},
  number={4},
  pages={457--477},
  year={1993},
  publisher={Elsevier}
}

@article{schoppe2013comparisons,
  title={Comparisons of levels and predictors of mothers' and fathers' engagement with their preschool-aged children},
  author={Schoppe-Sullivan, Sarah J and Kotila, Letitia E and Jia, Rongfang and Lang, Sarah N and Bower, Daniel J},
  journal={Early child development and care},
  volume={183},
  number={3-4},
  pages={498--514},
  year={2013},
  publisher={Taylor \& Francis}
}

@article{hernawati2020differences,
  title={Differences in father and mother involvement and the factors that influence it on early childhood education},
  author={Hernawati, Neti and Herawati, Tin},
  journal={SEA-CECCEP},
  volume={1},
  number={01},
  year={2020}
}

@article{braun2006using,
  title={Using thematic analysis in psychology},
  author={Braun, Virginia and Clarke, Victoria},
  journal={Qualitative research in psychology},
  volume={3},
  number={2},
  pages={77--101},
  year={2006},
  publisher={Taylor \& Francis}
}

@article{braun2019reflecting,
  title={Reflecting on reflexive thematic analysis},
  author={Braun, Virginia and Clarke, Victoria},
  journal={Qualitative research in sport, exercise and health},
  volume={11},
  number={4},
  pages={589--597},
  year={2019},
  publisher={Taylor \& Francis}
}

@article{terry2017thematic,
  title={Thematic analysis},
  author={Terry, Gareth and Hayfield, Nikki and Clarke, Victoria and Braun, Virginia and others},
  journal={The SAGE handbook of qualitative research in psychology},
  volume={2},
  number={17-37},
  pages={25},
  year={2017},
  publisher={SAGE Publications Ltd}
}

@article{xu2023superclue,
  title={Superclue: A comprehensive chinese large language model benchmark},
  author={Xu, Liang and Li, Anqi and Zhu, Lei and Xue, Hang and Zhu, Changtai and Zhao, Kangkang and He, Haonan and Zhang, Xuanwei and Kang, Qiyue and Lan, Zhenzhong},
  journal={arXiv preprint arXiv:2307.15020},
  year={2023}
}

@misc{MOE_3to6_ELDG_2012,
  author       = {Ministry of Education of the People’s Republic of China},
  title        = {Early Learning and Development Guidelines for Children Aged 3 to 6 Years},
  howpublished = {Online: PDF},
  year         = {2012},
  note         = {Accessed: 2025-11-26},
  url          = {https://www.unicef.cn/sites/unicef.org.china/files/2018-10/2012-national-early-learning-development-guidelines.pdf}
}

@article{channell2013individual,
  title={Individual differences in preschoolers’ emotion content memory: The role of emotion knowledge},
  author={Channell, Marie Moore and Barth, Joan M},
  journal={Journal of experimental child psychology},
  volume={115},
  number={3},
  pages={552--561},
  year={2013},
  publisher={Elsevier}
}

@article{liwag1995children,
  title={Children's memory for emotional events: The importance of emotion-related retrieval cues},
  author={Liwag, Maria D and Stein, Nancy L},
  journal={Journal of experimental child psychology},
  volume={60},
  number={1},
  pages={2--31},
  year={1995},
  publisher={Elsevier}
}

@article{hardy2020using,
  title={Using positive reinforcement with young children},
  author={Hardy, Jessica K and McLeod, Ragan H},
  journal={Beyond Behavior},
  volume={29},
  number={2},
  pages={95--107},
  year={2020},
  publisher={SAGE Publications Sage CA: Los Angeles, CA}
}

@article{sigler2005positive,
  title={From positive reinforcement to positive behaviors: An everyday guide for the practitioner},
  author={Sigler, Ellen A and Aamidor, Shirley},
  journal={Early Childhood Education Journal},
  volume={32},
  number={4},
  pages={249--253},
  year={2005},
  publisher={Springer}
}

@inproceedings{li2025you,
  title={“You Look Yummy!”: Designing a Tangible Interactive Game to Support Cognitive Development in Children with Low-Functioning Autism},
  author={Li, Jiajia and Chai, Yaqing and Wang, Cheng and Zhao, Jingyi and Liu, Shuqin and Duan, Ruiguang},
  booktitle={Adjunct Proceedings of the 38th Annual ACM Symposium on User Interface Software and Technology},
  pages={1--3},
  year={2025}
}

@inproceedings{cai2023starrypia,
  title={Starrypia: An AR gamified music adjuvant treatment application for children with autism based on combined therapy},
  author={Cai, Yu and Liu, Zhao and Yang, Zhuo and Tan, Yilan and Zhang, Junwei and Tang, Shuo},
  booktitle={Proceedings of the 36th Annual ACM Symposium on User Interface Software and Technology},
  pages={1--16},
  year={2023}
}

@inproceedings{li2021faceme,
  title={FaceMe: an augmented reality social agent game for facilitating children's learning about emotional expressions},
  author={Li, Jiajia and Zheng, Zixia and Wei, Xiemin and Wang, Guanyun},
  booktitle={Adjunct Proceedings of the 34th Annual ACM Symposium on User Interface Software and Technology},
  pages={17--19},
  year={2021}
}

@inproceedings{chen2025scenic,
  title={SCENIC: A Location-based System to Foster Cognitive Development in Children During Car Rides},
  author={Chen, Liuqing and Song, Yaxuan and Lyu, Ke and Xiao, Shuhong and Shen, Yilang and Sun, Lingyun},
  booktitle={Proceedings of the 38th Annual ACM Symposium on User Interface Software and Technology},
  pages={1--18},
  year={2025}
}

@inproceedings{guo2023sparkybot,
  title={Sparkybot: An Embodied AI Agent-Powered Robot with Customizable Characters andInteraction Behavior for Children},
  author={Guo, Yijie and Huang, Zhenhan and Wang, Ruhan and Li, Chih-Heng and Wu, Ruoyu and Sun, Qirui and Yao, Zhihao and Mi, Haipeng and Peng, Yu},
  booktitle={Adjunct Proceedings of the 36th Annual ACM Symposium on User Interface Software and Technology},
  pages={1--3},
  year={2023}
}

@inproceedings{zhang2025empowering,
  title={Empowering Children to Create AI-Enabled Augmented Reality Experiences},
  author={Zhang, Lei and Zhou, Shuyao and Liaqat, Amna and Mak, Tinney and Berengard, Brian and Qian, Emily and Monroy-Hern{\'a}ndez, Andr{\'e}s},
  booktitle={Proceedings of the 38th Annual ACM Symposium on User Interface Software and Technology},
  pages={1--16},
  year={2025}
}

@inproceedings{zhao2023narratron,
  title={Narratron: Collaborative writing and shadow-playing of children stories with large language models},
  author={Zhao, Yubo and Bao, Xiying},
  booktitle={Adjunct Proceedings of the 36th Annual ACM Symposium on User Interface Software and Technology},
  pages={1--6},
  year={2023}
}

@inproceedings{shi2023operar,
  title={OperAR: Using an augmented reality agent to enhance children's interactive intangible cultural heritage experience of the Peking Opera},
  author={Shi, Hongning and Li, Jiajia and Xue, Lian and Song, Yajing},
  booktitle={Adjunct Proceedings of the 36th Annual ACM Symposium on User Interface Software and Technology},
  pages={1--3},
  year={2023}
}

@inproceedings{vargas2023talemate,
  title={TaleMate: Collaborating with Voice Agents for Parent-Child Joint Reading Experiences},
  author={Vargas-Diaz, Daniel and Karunaratna, Sulakna and Kim, Jisun and Lee, Sang Won and Choi, Koeun},
  booktitle={Adjunct Proceedings of the 36th Annual ACM Symposium on User Interface Software and Technology},
  pages={1--3},
  year={2023}
}

@inproceedings{higuchi2018visualizing,
  title={Visualizing gaze direction to support video coding of social attention for children with autism spectrum disorder},
  author={Higuchi, Keita and Matsuda, Soichiro and Kamikubo, Rie and Enomoto, Takuya and Sugano, Yusuke and Yamamoto, Junichi and Sato, Yoichi},
  booktitle={Proceedings of the 23rd International Conference on Intelligent User Interfaces},
  pages={571--582},
  year={2018}
}

@inproceedings{song2016talklime,
  title={TalkLIME: mobile system intervention to improve parent-child interaction for children with language delay},
  author={Song, Seokwoo and Kim, Seungho and Kim, John and Park, Wonjeong and Yim, Dongsun},
  booktitle={Proceedings of the 2016 ACM International Joint Conference on Pervasive and Ubiquitous Computing},
  pages={304--315},
  year={2016}
}

@incollection{sohail2025rehnuma,
  title={Rehnuma: Interactive Companions for Speech-Delayed Children},
  author={Sohail, Nimra and Zaidi, Samiya Ali and Yasin, Huzaifah and Bhatti, Neelma and Shahid, Suleman},
  booktitle={Proceedings of the 24th Interaction Design and Children},
  pages={938--943},
  year={2025}
}

@inproceedings{leonidis2017home,
  title={Home game: an educational game for children with cognitive impairments},
  author={Leonidis, Asterios and Arampatzis, Dimitris and Korozi, Maria and Adami, Ilia and Ntoa, Stavroula and Stephanidis, Constantine},
  booktitle={Proceedings of the 2017 Conference on Interaction Design and Children},
  pages={667--670},
  year={2017}
}

@inproceedings{washington2016wearable,
  title={A wearable social interaction aid for children with autism},
  author={Washington, Peter and Voss, Catalin and Haber, Nick and Tanaka, Serena and Daniels, Jena and Feinstein, Carl and Winograd, Terry and Wall, Dennis},
  booktitle={Proceedings of the 2016 CHI Conference Extended Abstracts on Human Factors in Computing Systems},
  pages={2348--2354},
  year={2016}
}

% it can be deleted if there's no appendix
\clearpage
\appendix
\newpage

\section{PACEE System Implementation}
\label{appendix:PACEE System Implementation}
We implemented the core agent workflow in Python on a Flask server. The PACEE interface, written in Vue, interacts with the server by exchanging messages via WebSocket. Children's emotional profile and conversation graph information are stored in separate JSON files. 

For most text generation tasks, we leveraged the GPT-4o API to balance speed and output quality. To minimize latency in agent conversations and real-time advice delivery, the faster GPT-4o-mini model was adopted. Speech-to-text processing utilized the TINGWU API, with audio sliced and compressed on the frontend before transcription. Text-to-speech conversion relied on Alibaba Cloud's streaming synthesis API, while image generation integrated the Stable Diffusion XL (SDXL) model.

\section{Additional Formative Study Details}
\label{appendix:Additional Formative Study Details}
This section provides supplementary details about the formative study, including (1) participant demographics (\autoref{tab:formative-study-demographic-1-1}-\autoref{tab:formative-study-demographic-2}) and (2) the scenario database identified from the study, which captures common situations where children aged 3-6 experience negative emotions and require repeated parental guidance (\autoref{tab:scenario-database}).

\begin{table}[h]  
\caption{Demographic details of kindergarten teachers in the need-finding interviews.}
\centering  
\small
\begin{tabular}{lll}
\toprule[1pt]
\multicolumn{1}{c}{\textbf{Alias}} & \multicolumn{1}{c}{\textbf{Age}} & \multicolumn{1}{c}{\textbf{Gender}} \\ \hline
\textbf{K1}                        & 31                               & Male                                \\ \hline
\textbf{K2}                        & 29                               & Female                              \\ \hline
\textbf{K3}                        & 30                               & Male                                \\ \hline
\textbf{K4}                        & 48                               & Female                              \\ \hline
\textbf{K5}                        & 31                               & Female                              \\ \bottomrule[1pt]
\end{tabular}
\label{tab:formative-study-demographic-1-1}
\end{table}

\begin{table}[h]  
\caption{Demographic details of parents in the need-finding interviews.}
\centering  
\small
\begin{tabular}{lll}
\toprule[1pt]
\multicolumn{1}{c}{\textbf{Alias}} & \multicolumn{1}{c}{\textbf{Age}} & \multicolumn{1}{c}{\textbf{Gender}} \\ \hline
\textbf{G1}                        & 37                               & Female                              \\ \hline
\textbf{G2}                        & 45                               & Female                              \\ \hline
\textbf{G3}                        & 41                               & Female                              \\ \hline
\textbf{G4}                        & 38                               & Male                                \\ \hline
\textbf{G5}                        & 44                               & Female                              \\ \bottomrule[1pt]
\end{tabular}
\label{tab:formative-study-demographic-1-2}
\end{table}

\begin{table}[h]  
\caption{Demographic details of participants in the observational experiments.}  
\centering  
\small
\begin{tabular}{lllll}
\toprule[1pt]
\textbf{Alias} & \multicolumn{2}{l}{\textbf{Parent}} & \multicolumn{2}{l}{\textbf{Child}} \\ \hline
\textbf{}      & Age             & Gender            & Age            & Gender            \\ \hline
\textbf{O1}    & 37              & Male              & 3              & Boy               \\ \hline
\textbf{O2}    & 37              & Female            & 6              & Boy               \\ \hline
\textbf{O3}    & 41              & Female            & 4              & Boy               \\ \hline
\textbf{O4}    & 43              & Female            & 6              & Boy               \\ \hline
\textbf{O5}    & 33              & Female            & 5              & Girl              \\ \bottomrule[1pt]
\end{tabular} 
\label{tab:formative-study-demographic-2}
\end{table}

\begin{table*}[htbp]
\centering
\caption{Our need-finding interviews identified common scenarios of negative emotions experienced by children aged 3-6.}
\small
\begin{tabular}{ccc}
\toprule[1pt]
\textbf{Type}        & \textbf{Example Scenarios}                                                                                          & \textbf{Common Emotions} \\ \hline
Separation           & \begin{tabular}[c]{@{}c@{}}Leaving parents for school; \\ missing parents; waking up alone\end{tabular}             & Anxiety, sadness         \\ \hline
Peer conflict        & \begin{tabular}[c]{@{}c@{}}Disputes over toys/turns; sibling \\ quarrels; having belongings taken\end{tabular}      & Anger, frustration       \\ \hline
Social setbacks      & \begin{tabular}[c]{@{}c@{}}Being ignored by peers; \\ unable to join new groups\end{tabular}                        & Loneliness, sadness      \\ \hline
Physical discomfort  & \begin{tabular}[c]{@{}c@{}}Injury, illness, falls, or \\ grogginess upon waking\end{tabular}                        & Distress, sadness        \\ \hline
Autonomy violation   & \begin{tabular}[c]{@{}c@{}}Forced task termination (e.g., iPad \\ taken away); not winning first place\end{tabular} & Anger, disappointment    \\ \hline
Negative feedback    & \begin{tabular}[c]{@{}c@{}}Teacher corrections; \\ critical remarks from peers\end{tabular}                         & Sadness, shame           \\ \hline
Stressful challenges & \begin{tabular}[c]{@{}c@{}}Challenging sports/games; \\ performing on stage\end{tabular}                            & Anxiety, fear            \\ \bottomrule[1pt]
\end{tabular}
\label{tab:scenario-database}
\end{table*}

\section{Additional User Evaluation Details}
\label{appendix:Additional User Evaluation Details}
This section provides supplementary details about the user evaluation, including (1) participant demographics (\autoref{tab:evaluation-demographic}) and (2) an analysis of real-time advice accepted by parents during the study (\autoref{tab:accepted_advice}).

\begin{table*}[htb]
\caption{Participant demographic information in the user evaluation.}
\small
\begin{tabular}{cccccc}
\toprule[1pt]
\textbf{Child ID} & \textbf{Child Gender} & \textbf{Child Age} & \textbf{Parent ID} & \textbf{Parent Gender} & \textbf{Parent Age} \\ \hline
C1                & Female                & 5                  & P1                 & Male                   & 41                  \\ \hline
C2                & Male                  & 5                  & P2                 & Male                   & 36                  \\ \hline
C3                & Female                & 4                  & P3                 & Female                 & 37                  \\ \hline
C4                & Male                  & 5                  & P4                 & Female                 & 43                  \\ \hline
C5                & Female                & 5                  & P5                 & Female                 & 38                  \\ \hline
C6                & Female                & 4                  & P6                 & Female                 & 43                  \\ \hline
C7                & Female                & 5                  & P7                 & Female                 & 34                  \\ \hline
C8                & Male                  & 6                  & P8                 & Female                 & 36                  \\ \hline
C9                & Female                & 5                  & P9                 & Female                 & 39                  \\ \hline
C10               & Female                & 4                  & P10                & Female                 & 29                  \\ \hline
C11               & Male                  & 5                  & P11                & Female                 & 39                  \\ \hline
C12               & Male                  & 5                  & P12                & Female                 & 39                  \\ \hline
C13               & Male                  & 6                  & P13                & Male                   & 39                  \\ \hline
C14               & Male                  & 5                  & P14                & Female                 & 36                  \\ \hline
C15               & Male                  & 6                  & P15                & Male                   & 36                  \\ \hline
C16               & Male                  & 6                  & P16                & Female                 & 38                  \\ \bottomrule[1pt]
\end{tabular}
\label{tab:evaluation-demographic}
\end{table*}

\begin{table*}[htbp]
\caption{Categories of real-Time advice accepted by parents during the user evaluation.}
\small
\begin{tabular}{ccll}
\toprule[1pt]
\textbf{Type}                                                            & \textbf{Number} & \multicolumn{1}{c}{\textbf{Description}}                                                                                                                & \multicolumn{1}{c}{\textbf{Example}}                                                                                                                      \\ \hline
\begin{tabular}[c]{@{}c@{}}Open-ended \\ Questioning\end{tabular}        & 22              & \begin{tabular}[c]{@{}l@{}}Asking open-ended questions to \\ encourage children to describe their \\ emotions, thoughts, or experiences.\end{tabular}   & \begin{tabular}[c]{@{}l@{}} \textit{Do you have any special feelings} \\ \textit{when redeeming your points? Do} \\ \textit{you feel happy or a bit unsure?} \end{tabular}             \\ \hline
\begin{tabular}[c]{@{}c@{}}Scenario \\ Simulation\end{tabular}           & 16              & \begin{tabular}[c]{@{}l@{}}Setting up or recalling specific \\ scenarios to encourage role-playing \\ or imagination.\end{tabular}                      & \begin{tabular}[c]{@{}l@{}} \textit{Imagine yourself as a brave little} \\ \textit{hero, like My Little Pony---you can} \\ \textit{show your courage and talent on stage.}\end{tabular} \\ \hline
\begin{tabular}[c]{@{}c@{}}Concrete \\ Suggestions\end{tabular}          & 7               & \begin{tabular}[c]{@{}l@{}}Providing clear, specific strategies \\ to help children manage emotions \\ or enhance positive feelings.\end{tabular}       & \begin{tabular}[c]{@{}l@{}} \textit{We can go to the small playground} \\ \textit{on the rooftop for a run, or maybe} \\ \textit{find something delicious to eat.}\end{tabular}       \\ \hline
\begin{tabular}[c]{@{}c@{}}Positive \\ Encouragement\end{tabular}        & 9               & \begin{tabular}[c]{@{}l@{}}Using praise and positive feedback \\ to reinforce children's positive \\ behaviors and emotional experiences.\end{tabular}  & \begin{tabular}[c]{@{}l@{}} \textit{You did a great job jumping rope---}\\ \textit{210 jumps is really impressive!} \end{tabular}                                               \\ \hline
\begin{tabular}[c]{@{}c@{}}Empathy and \\ Acceptance\end{tabular}        & 16              & \begin{tabular}[c]{@{}l@{}}Expressing empathy and acceptance \\ towards children's emotions, making \\ them feel understood and supported.\end{tabular} & \begin{tabular}[c]{@{}l@{}} \textit{I understand you might feel nervous} \\ \textit{on such a big stage. Dad has felt like} \\ \textit{that before, too.}\end{tabular}                \\ \hline
\begin{tabular}[c]{@{}c@{}}Collaborative \\ Problem-solving\end{tabular} & 31              & \begin{tabular}[c]{@{}l@{}}Discussing solutions together with \\ children.\end{tabular}                                                                 & \begin{tabular}[c]{@{}l@{}}\textit{Let's think together about what we can} \\ \textit{do to perform even better in the next} \\ \textit{competition.} \end{tabular}                   \\ \bottomrule[1pt]
\end{tabular}
\label{tab:accepted_advice}
\end{table*}

%TC:endignore
\end{document}